\documentclass[twocolumn,preprintnumbers,a4paper,superscriptaddress,
aps,showpacs,floatfix]{revtex4}
\usepackage{graphicx}
\usepackage{bm}
\begin{document}
\def\la{\langle}
\def\ra{\rangle}
\newcommand{\beq}{\begin{equation}}
\newcommand{\eeq}{\end{equation}}
\newcommand{\beqa}{\begin{eqnarray}}
\newcommand{\eeqa}{\end{eqnarray}}
\newcommand{\intf}{\int_{-\infty}^\infty}
\newcommand{\into}{\int_0^\infty}

\title{Suppression of Rabi oscillations for moving atoms}
\author{B. Navarro}
\affiliation{Departamento de Qu\'\i mica-F\'\i sica, Universidad del
Pa\'\i s Vasco, Apdo. 644, 48080 Bilbao, Spain}
\affiliation{Fisika Teorikoaren Saila, Euskal Herriko Unibertsitatea, 
644 P. K., 48080 Bilbao, Spain}
\author{I. L. Egusquiza}
\affiliation{Fisika Teorikoaren Saila, Euskal Herriko Unibertsitatea, 
644 P. K., 48080 Bilbao, Spain}
\author{J. G. Muga}
\affiliation{Departamento de Qu\'\i mica-F\'\i sica, Universidad del
Pa\'\i s Vasco, Apdo. 644, 48080 Bilbao, Spain}
\author{G. C. Hegerfeldt}
\affiliation{Institut f\"ur Theoretische Physik,  Universit\"at G\"ottingen,
 37073 G\"ottingen, Germany}

%\author{......}
%\affiliation{Departamento de Qu\'\i mica-F\'\i sica, Universidad del
%Pa\'\i s Vasco, Apdo. 644, 48080 Bilbao, Spain}

\begin{abstract}
The well-known laser-induced Rabi oscillations of a two-level atom are
shown to be suppressed under certain conditions  
when the atom is entering a laser-illuminated region. 
For temporal Rabi oscillations the effect has two regimes:
classical-like, at intermediate atomic velocities, 
and quantum at low velocities, associated 
respectively with the formation of incoherent or coherent
internal states of the atom in the laser region.
In the low velocity regime 
the laser projects the atom onto a pure internal state
that can be controlled by detuning.
Spatial Rabi oscillations are only suppressed in this 
low velocity, quantum regime.  

\end{abstract}
\pacs{03.65.-w, 32.80.-t, 42.50.-p}
\maketitle

\section{Introduction}
The coupling between the center-of-mass motion of the atom 
and localized laser fields is a central topic in quantum optics, and
leads to many 
important effects and applications in cooling, trapping,  deflection,
and isotope separation  
experiments. The objective of this 
paper is to point out one further consequence of such a 
coupling: the suppression of Rabi oscillations and the related 
possibility of controlled projection  
onto internal pure states by the motion of the atom from
a laser-free region into a laser-illuminated region.  

For a general two-level system with an interaction-picture 
Hamiltonian of the form 
\beq
H= \frac{\hbar}{2} \left({0\atop \Omega}{\Omega
    \atop -2\delta} \right),
\label{ham}
\eeq
a state ${\bf \Psi}\equiv{\Psi^{(1)}\choose\Psi^{(2)}}$, beginning in the
ground state $|1\ra={1\choose 0}$ at $t=0$, will evolve  
according to 
\begin{eqnarray}
\Psi^{(1)}&=&
e^{i\delta t/2}\left[\cos\left(\frac{\Omega't}{2}\right)
-i\frac{\delta}{\Omega'}\sin\left(\frac{\Omega't}{2}\right)\right],
\nonumber\\ 
\Psi^{(2)}&=&  
\frac{-ie^{i\delta t/2}\Omega}{\Omega'}\sin\left(\frac{\Omega' t}{2}\right), 
\label{eq110}
\end{eqnarray}
where 
\beq
\Omega'=(\Omega^2+\delta^2)^{1/2}
\eeq
and $\Omega$ is the usual Rabi frequency. 
%Note that the excited state population $P_2$ oscillates between
%0 and 1 for $\delta=0$, and between 0 and $(\Omega/\Omega')^2$ 
%for an arbitrary $\delta$.  

There is currently much interest in the dynamics of
two-level systems 
governed by the Hamiltonian of Eq. (\ref{ham}) 
plus an additional time-dependent driving term,
in particular to determine the Rabi oscillation
suppression 
for critical values of the external field parameters.  
%These may be associated, for example, with tunneling 
%between two wells representing stable configurations of a
%molecular system.
In the so called
dynamical localization effect, for example,
the system remains in one of the two levels \cite{Sacchetti01}.
Another Rabi-oscillation-suppression effect has been described recently for  
the scattering of atoms by
a standing light wave \cite{EFYS02}. 
In this case, the partial contributions from different 
diffraction angles lead to
different oscillation periods and thus to 
inhomogeneous broadening.  

In this paper we describe a different type of oscillation
suppression due to the  motion of the system rather
than to an additional time dependent field, or to a diffraction 
effect.
%Moreover, the system will end up in a mixture (coherent or incoherent) 
%of the two levels  
%instead of staying in only one of them.  
Explicitly, we shall consider the motion of a two-level atom with
transition frequency $\omega$
in one dimension, with a classical electric field (laser of frequency
$\omega_L$) illuminating the half axis $x>0$ perpendicularly. Without
damping,  and in an interaction picture for the internal degrees of
freedom, the Hamiltonian  can be written in the form 
\begin{equation}
H= \hat{p}^2/2m + \frac{\hbar}{2} \left({0\atop \Omega\Theta(\hat{x})}
{\Omega\Theta(\hat{x})
    \atop -2\delta} \right),
\label{H}
\end{equation}
where $\Theta$ is the step function, $\Omega$ plays the role of
a laser-atom  coupling constant, $\delta = \omega_L -\omega$ is
the detuning, and hats denote operators  
whenever confusion is possible with a corresponding $c$-number.
In particular, 
$\hat{p}$ is the momentum operator for the $x$ direction. 
If one takes damping due to photon emissions into account, it can be shown,
by means of the quantum jump approach \cite{Hegerfeldt91}, that in
three dimensions the
atomic time development between emissions  may be described by an
(effective, non-hermitian) ``conditional'' Hamiltonian $H_{\rm c}$,
\beqa\nonumber
H_{\rm c} &=& {\bf{\hat p}}^2/2m + \frac{\hbar}{2} \Omega\, \Theta (\hat{x})
\left\{ |2\rangle\langle1| e^{i k_L \hat{y}}+
{\rm h.c.}\right\}
\\
\label{2.4}
&-&\frac{\hbar}{2}(2\delta+i\gamma)|2\rangle\langle2|,
\eeqa
where $\gamma$ is the Einstein 
coefficient of level 2, i.e. its decay rate or inverse life time,
and ${\bf{\hat p}}$ is the 
momentum operator in three dimensions (3D). The  factor
$e^{i k_L \hat{y}}$
takes into account  the spatial dependence of the laser coupling.
A Hamiltonian of the form of Eq. (\ref{H}) is obtained from
Eq. (\ref{2.4}) by neglecting
spontaneous emissions, i.e. setting $\gamma=0$, and   
assuming in addition that the atomic
wave packet is centered at $y=0$ and satisfies $k_L\Delta y\ll 1$, 
at least for some time,  so that the exponentials can be dropped 
and a one dimensional kinetic term  suffices. This approximation and its 
limitations will be further  commented on Appendix \ref{limits}.

The conditional Hamiltonian $H_{\rm c}$ is closely related to
waiting times between photon emissions \cite{Cohen-Dalibard85}. Indeed,
let an atomic state $|\Psi(0)\rangle$ be prepared at time $0$. Then it
can be shown by 
means of the quantum jump approach \cite{Hegerfeldt91} that  
\begin{equation}\label{2.3}
P_0(t)\equiv 
||e^{-iH_{\rm c}(t)/\hbar} |\Psi(0)\rangle||^2
\end{equation}
gives the probability of no emission until time $t$ and $\Pi(t)\equiv
-P_0'(t)$ is the probability density for the first photon. After an
emission the atom has to be reset, here to its ground state, and then
the conditional time development resumes. In this way one can simulate
an emission sequence and the corresponding Bloch (master) equation for
the atom.

\begin{figure}
{\includegraphics[width=3.35in,height=2.25in]{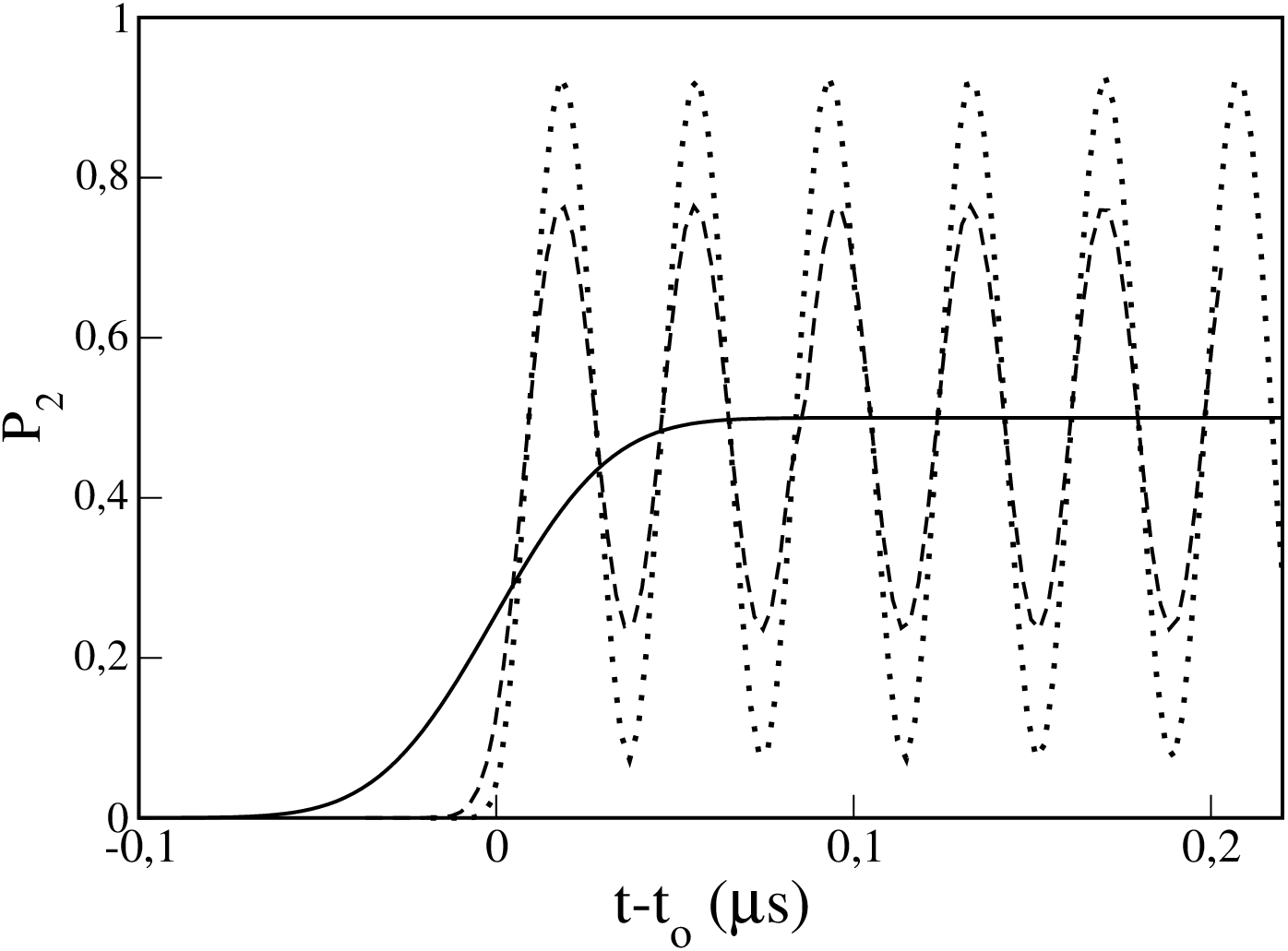}}
\caption{Probability of excited state $P_2$
versus $t-t_0$, where $t_0$ is a mean entrance time, see the
definition below.
$\Omega=166.5\times 10^6$ s$^{-1}$, $\gamma=\delta=0$. 
The initial states (at $t=0$) for the center of mass 
are minimum-uncertainty product Gaussians with $\Delta_x=0.24\,\mu$m,
$\la x\ra=-1.32\, \mu$m,
and $\la v\ra=9.03$ m/s (solid line), 
49.68 m/s (dotted line) and 
36.13 m/s (dashed line). $t_0=-\la x\ra/\la v\ra$.
The atomic mass is taken to be that of Cs here and in all 
following figures.\vspace*{.5cm}\\ 
}
\label{resumen}
\end{figure}

As we will show, for moving atoms there are Rabi oscillations both in
time as well as in space. In the temporal case  one asks for the
probability, $P_2(t)$, of 
finding the atom in the excited state, without regard to its spatial
position. For the simple model in Eq. (\ref{H}) without damping, this
can be easily calculated and  Figure \ref{resumen} depicts a
suppression for this case. The solid line  shows the population of the
excited state $P_2$ 
versus $t$ for $\delta=0$ and for a wave packet which is prepared
at $t=0$ with negligible negative
momentum components as a minimum-uncertainty-product
Gaussian, with the atom in the ground state, and far from the laser.
At sufficiently large times, and for the parameters we have chosen,
the wave packet  is completely within the laser region,   
but there are no Rabi oscillations.
Instead, $P_2$ increases monotonically, and saturates at $0.5$; 
this might look surprising at first sight: 
there is no classical averaging (the state is quantum and coherent),
there is no dispersion 
related broadening, there is no time dependent term in
the Hamiltonian,  
there is not even decay that could smooth or suppress
the oscillation. 
\begin{figure}
%$^{}$\vspace*{1.cm}\\ 
{\includegraphics[width=3.35in]{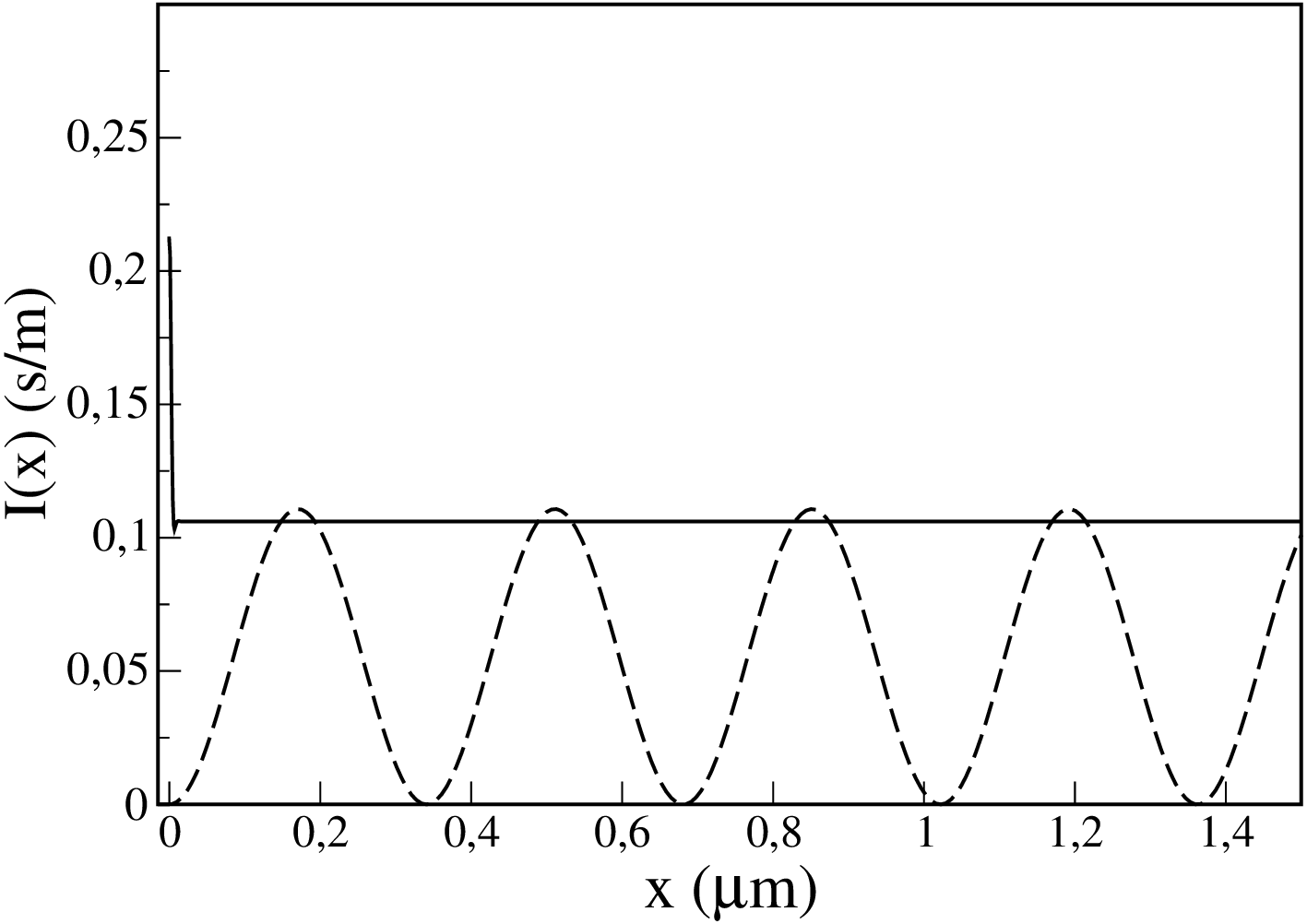}}
\caption{Probability density of the excited state, integrated over time,
for two different wave packets: $\la v\ra =0.0090$ m/s (solid line) and 
$\la v\ra=9.03$ m/s (dashed line); $\Omega=166.5\times 10^6$ s$^{-1}$;
$\gamma=\delta=0$. The initial state is a minimum-uncertainty-product Gaussian 
in the ground state with 
$\Delta_x=0.24\, \mu$m, and $\la x\ra=-1.32\mu$m. With these parameters  
$v_c=k_c\hbar/m=0.28$ m/s and $v_R=k_R\hbar/m=32.3$ m/s,
see Eqs. (\ref{kc}) and
(\ref{esti}).\vspace*{.3cm}\\ 
}  
\label{px015}
\end{figure} 
 Basically, for measurements in time 
the effect has two
different regimes and explanations depending on the incident energy:
at very low kinetic energy, the suppression is due to the formation of a pure 
non-oscillating internal state of the atom in the laser region, 
whereas at intermediate energies
it may be understood in terms of a simple semiclassical
approximation. Yet at higher energies the effect disappears, 
and the expected Rabi oscillations may be seen more and more clearly 
for increasing velocities, as illustrated by the dashed and 
dotted lines of Figure \ref{resumen}.

Spatial Rabi oscillations can  occur in the
emission probability at different atomic positions. If 
$P_2(x,t)$ denotes the probability density to find the
atom in its excited state at the position $x$ at time $t$ and if one
defines $I(x)$ by
%then $\gamma
%P_2(x,t)$ is, when all atoms emit photon,
%the joint probability density to find a spontaneous photon
%emission at $t$ and  the atom at $x$. Hence, considering a stream of
%identically prepared atoms and defining
%
\beq
I(x)\equiv \int dt P_2(x,t),
\label{taux}
\eeq
then $\gamma I(x)dx$ gives the
number of emitted photons (per atom) with the atom at $dx$, 
and thus $I(x)$ is proportional to the
photon intensity for the atom at position $x$ \cite{foot1}.
%{$I(x)$ 
%may be also 
%interpreted as a dwell time density for the
%excited state at $x$ \cite{tqm}}. 
Neglecting damping, the expression on the rhs can be calculated with
Eq. (\ref{H}), and an example is given in Fig. \ref{px015}, where the 
spatial Rabi oscillation is completely absent in one of the two cases 
depicted.  
In contrast to the temporal case,  spatial oscillations 
are only suppressed by the quantum mechanism, i.e., at very low
kinetic energies.

In the next four sections we shall characterize, 
neglecting damping due to spontaneous emissions,
several aspects of the Rabi oscillation
suppression due to atomic motion: 
the basic theory (Section \ref{fnd}), the effect of the velocity
regimes in the 
Rabi oscillations for measurements in time (Section \ref{evros}),
the effect of detuning 
and the velocity in the internal atomic states (Section \ref{evdis}),
and the suppression of spatial Rabi oscillations (Section \ref{rospd}).
The effect of damping will be 
taken into account in Section  \ref{sd}.

\section{Basic theory\label{fnd}}
\subsection{Exact results}
Here we shall present the theory required to describe 
the interaction of the moving atom with the perpendicular laser. We
first neglect damping and 
solve  the eigenvalue equation for the Hamiltonian of
Eq. (\ref{H}) subject to the condition
that the atom impinges on the laser beam from the left in the ground state.  
The eigenvalues are $E =\hbar^2k^2/2m$ and the eigenfunctions
are denoted by 
${\bf \Phi}_k\equiv{\phi^{(1)}\choose\phi^{(2)}}$,
with $k>0$. 
% $ {\bf \Phi}_k (x)$ have the form
%
%
%
%
%where $R_{1,2}$ are nonzero reflection amplitude coefficients and
%Im~$k_{\pm}\ge 0$. The explicit expressions for 

For $x \le 0$,
${\bf \Phi}_k$ is of the form
\begin{equation}\label{A15}
{\bf \Phi}_k (x) = \frac{1}{\sqrt{2\pi}} \left(  
{e^{ikx}+ R_1e^{-ikx} \atop R_2 e^{-iqx}}
 \right), \quad x\le0, \quad k>0, 
\end{equation}
where 
\beq
q^2=k^2+2m\delta/\hbar
\eeq
and ${\rm{Im}}(q)\ge 0$. For large enough (negative) detuning, 
$k<|2m\delta/\hbar|^{1/2}$, $q$ 
becomes purely imaginary and the 
reflected wave for the excited state decays exponentially. 

In the laser region,  
let $|\lambda_+\rangle$ and 
$|\lambda_-\rangle$ be the eigenstates of the
matrix $\frac{1}{2} \left( {0\atop \Omega}{\Omega\atop -2\delta}
\right)$ corresponding to the eigenvalues $\lambda_\pm$. One easily finds 
\begin{eqnarray}\label{A13}
\lambda_\pm &=& -\frac{1}{2}[\delta\pm \Omega'],
\\ \label{A14}
|\lambda_\pm \rangle &=&  {1 \choose \frac{-(\delta\pm\Omega')}{\Omega}}. 
\end{eqnarray} 
Note that $|\lambda_\pm \rangle$  
have not been normalized.

For $x \ge 0$, one can write ${\bf \Phi}_k$  as a superposition
of $|\lambda_\pm \rangle$, 
\begin{equation} \label{A17}
\sqrt{2\pi} {\bf \Phi}_k (x) = C_+ |\lambda_+ \rangle e^{ik_+x} +
C_- |\lambda_- \rangle e^{ik_-x}. 
\end{equation}
[The mathematical solutions $e^{-ik_\pm x}$ are not included 
because they correspond to negative momenta or
increasing exponentials.] 

From the eigenvalue equation $H {\bf \Phi}_k = E_k {\bf
\Phi}_k$  it follows that 
\begin{equation}\nonumber
k_\pm^2 = k^2 - 2m \lambda_\pm /\hbar = k^2 
+\frac{m}{\hbar}(\delta\pm\Omega'),
\end{equation}
with Im $\, k_\pm \ge 0$, 
and from the continuity of ${\bf \Phi}_k(x)$ at $x = 0$ one obtains, with
$|1\rangle = {1 \choose 0}$  and $ |2\rangle = {0 \choose 1}$,
\begin{eqnarray}\nonumber
1 + R_1 &=& C_+ \langle 1|\lambda_+ \rangle + C_- \langle 1|\lambda_-
\rangle \\ \nonumber
R_2 &=& C_+ \langle2| \lambda_+\rangle + C_- \langle 2|\lambda_- \rangle.
\end{eqnarray}
Similar relations result from the continuity of ${\bf \Phi}_k'(x)$
at $x = 0$, yielding
\begin{eqnarray}
C_+ &=& -2k(q+k_-)\lambda_-/D,
\nonumber
\\
\nonumber
C_- &=&2k(q+k_+)\lambda_+/D,
\\
\nonumber
R_1 &=& [\lambda_+(q+k_+)(k-k_-)-\lambda_-(q+k_-)(k-k_+)]/D,
\nonumber\\
R_2 &=&k(k_--k_+)\Omega/D,
\label{A18}
\end{eqnarray}
where 
\beq\label{d}
D=(k+k_-)(q+k_+)\lambda_+-(k+k_+)(q+k_-)\lambda_-.
\eeq
Thus Eq. (\ref{A17}) becomes, in components and for $x \ge 0$,
\begin{eqnarray}
\phi_k^{(1)}(x) &=&-\frac{2k}{\sqrt{2\pi} D}
\{(q+k_-)\lambda_- e^{ik_+x} - (q+k_+)\lambda_+e^{ik_-x}\},
\nonumber\\
\phi_k^{(2)}(x) &=&\frac{k\Omega}{\sqrt{2\pi} D}
\{(q+k_-)e^{ik_+x} - (q+k_+)e^{ik_-x}\}.
\end{eqnarray}
Time dependent wave packets incident on the laser region with positive 
momentum components and with the atom in the ground state 
may be formed by linear superposition \cite{DEHM02}, 
\beq
\label{tde}
{\bf \Psi}(x,t)=\int_0^\infty
dk\,e^{-i\hbar k^2t/2m}{\bf \Phi}_k(x) \tilde{\psi}(k),
\eeq
where $\tilde{\psi}(k)$ is the wave number amplitude at time $t=0$ 
corresponding to the freely moving packet. 

The conditional Hamiltonian $H_c$ for the one-dimensional case with
damping can be treated in a similar way, and this is done in Appendix
\ref{decay}.

\subsection{Different approximation regimes\label{3dvr}}
%
%
%
%
%
%We shall see below that For $\delta=0$ there is a clear
%separation between three velocity regimes
%that can be smoothed with a non zero detuning. 
Depending on the incident velocity different approximations are applicable.  
For large velocities a semiclassical approximation is sufficient
whereas 
for slow atoms it is essential to retain the quantum nature
of the translational motion. 
%In addition,  
%the semiclassical regime may be subdivided into 
%two when the observation is time dependent. This subdivision, 
%however, will not be seen  when the observations are 
%position dependent.  
%
\subsubsection{Fast atoms: semiclassical approximation}
For  ``high'' kinetic energy,  $k^2\gg |m(\delta\pm\Omega')/\hbar|$, 
we may approximate the wave numbers in the wave function by 
\beqa
k_\pm&\approx& k+ \frac{m}{2\hbar k}[\delta\pm \Omega'],
\label{hek}
\\
\label{heq}
q&\approx& k+\frac{m\delta}{k\hbar},
\eeqa
and therefore, 
\begin{eqnarray}
\phi_k^{(1)}(x) &\approx&
\frac{e^{ikx}} {\sqrt{2\pi}}\bigg\{e^{i\delta t/2}
\bigg[\cos\left(\frac{m\Omega'x}{2\hbar k}\right)
\nonumber
\\
&-&i\frac{\delta}{\Omega'}\sin\left(\frac{m\Omega'x}{2\hbar k}\right)
\bigg]\bigg\},
\nonumber\\ 
\phi_k^{(2)}(x) &\approx&
\frac{e^{ikx}}{\sqrt{2\pi}}\bigg\{e^{i\delta t/2}
\bigg[\frac{-i\Omega}{\Omega'}\sin\left(\frac{m\Omega' x}
{2\hbar k}\right)
\bigg]\bigg\}.
\label{eq11}
\end{eqnarray}
[The relation with the Raman-Nath approximation is 
discussed in Appendix \ref{RN}.] 
Up to the normalization factor this reveals a  
``spatial Rabi oscillation'' that may be associated with
the temporal one by the change of variable 
$t=xm/k\hbar$, the
time that a classical particle with 
momentum $k\hbar$ needs to 
travel from the origin to $x$.  
With this change the terms in curly brackets  correspond 
exactly to the 
internal state amplitudes of the atom at rest, see Eq. (\ref{eq110}),
while the multiplying plane wave denotes an undisturbed 
center of mass motion with fixed momentum.

%In this regime and for $x\ge 0$ the densities take the form  
%
%\beqa
%|\phi_k^{(1)}(x)|^2&\approx& \frac{1}{2\pi}
%\bigg[\cos^2\left(\frac{m\Omega' x}{2\hbar k}\right)+\frac{\delta^2}
%{\Omega^2+\delta^2}
%\sin^2\left(\frac{m\Omega' x}{2\hbar k}\right)\bigg]
%\nonumber\\
%|\phi_k^{(2)}(x)|^2&\approx& \frac{1}{2\pi}\frac{\Omega^2}{\Omega^2+\delta^2}
%\sin^2\left(\frac{m\Omega' x}{2\hbar k}\right).
%\label{densi}
%\eeqa
%
Moreover in this regime reflection is  
negligible, 
\beq
R_1\approx \left(\frac{m\Omega}{4\hbar k^2}\right)^2,\;\;\;
R_2\approx\frac{m\Omega}{4\hbar k^2}.
\eeq
\begin{figure}
{\includegraphics[width=3.35in]{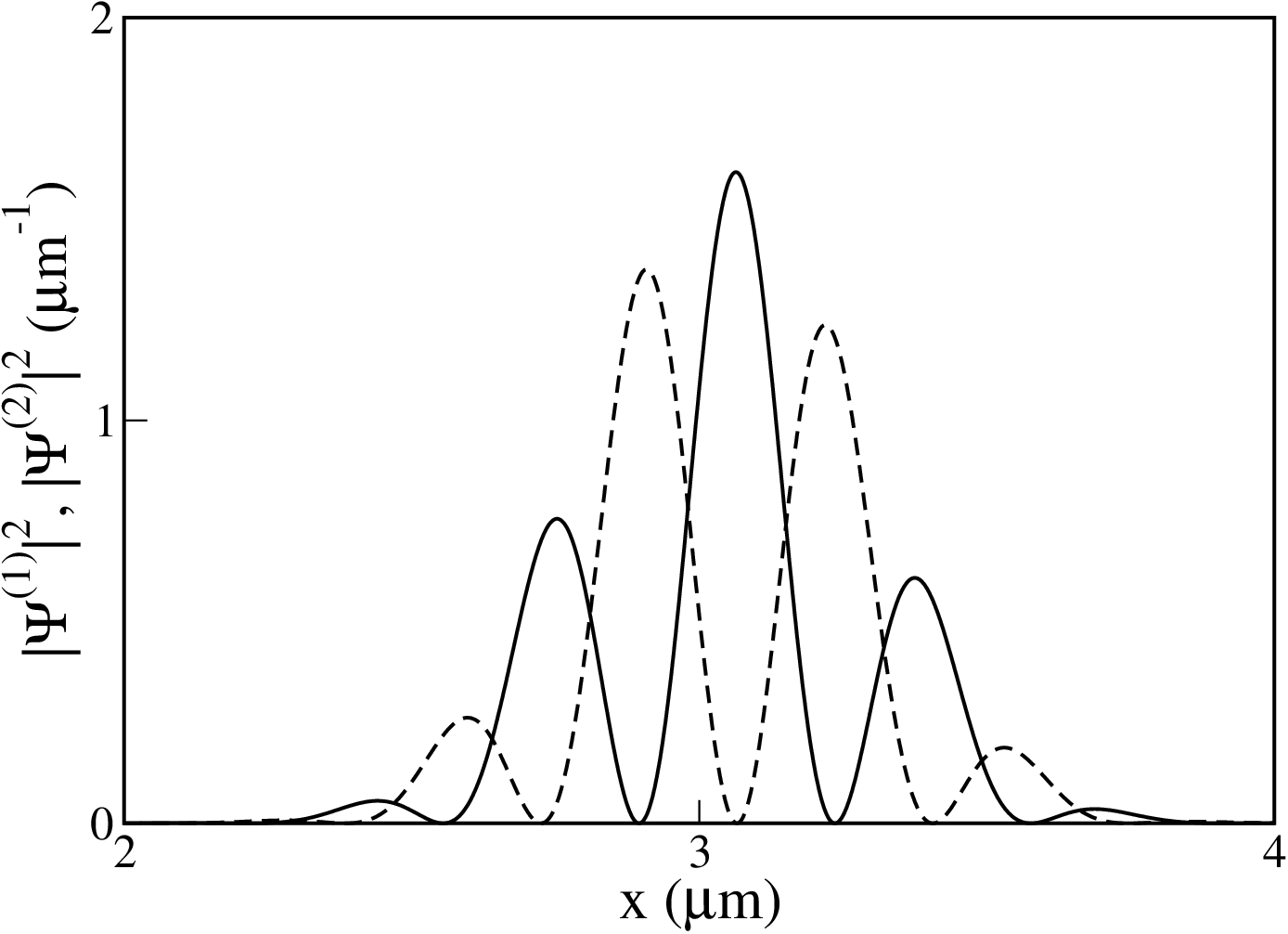}}
\caption{Densities for ground (solid line) and excited state (dashed line) 
at $t=0.5\, \mu$s. The initial 
state (at $t=0$) is 
a minimum-uncertainty product Gaussian for the center of mass, with  
$\Delta_x=0.2436\, \mu$m, $\la x\ra=-1.32\,\mu$m, 
and $\la v\ra =9.03$ m/s; $P_2(t=0)=0$; $\Omega=166.5 \times 10^6$
s$^{-1}$, $\gamma=\delta=0$, 
and the mass
corresponds to a Cs atom.}   
\label{densidades1}
\end{figure}
A further simplification is to assume that the 
interferences among different momenta are either negligible or average out.  
Then we may quite accurately approximate the center of mass motion of a 
wave packet with high kinetic energy 
by a purely classical free motion, 
whereas the internal state is treated quantum
mechanically. More specifically, in this 
approximation the wave packet width and spreading 
are taken into account by using an ensemble of semiclassical 
atoms with their center of mass distributed in 
classical phase space according to the
initial quantum Wigner function $W(q_0,p_0; t=0)$.
[For free motion, 
the quantum and ``classically evolved'' Wigner functions coincide at all times 
if initially identical.]
To each of these atoms we associate an internal state, that starts to 
oscillate once the center of mass crosses $x=0$ at $-q_0m/p_0$.  
The expectation values in the semiclassical approximation are
then computed by 
averaging over the ensemble of semiclassical atoms (translationally classical,
internally quantum). In particular,
the probability to find the atom in the excited state 
at a time $t$ is given by 
\beqa
P_2(t)&=&\intf dq_0\into dp_0\, W(q_0,p_0; t=0)
\\
&\times&\left(\frac{\Omega}{\Omega'}\right)^2
\sin^2[\Omega'(t+q_0m/p_0)/2]
\Theta(t+q_0m/p_0), 
\nonumber
\eeqa
where we have assumed that $t>0$, that the wave packet at $t=0$ is far from 
the laser, and that all atoms have positive momentum. 

\subsubsection{Slow atoms\label{sa}}

At the critical wavenumber $k=k_c$,
\beq\label{kc}
k_c\equiv [m(\Omega'-\delta)/\hbar]^{1/2},
\eeq
the de Broglie
wavelength for the atomic motion  $\lambda_{dB}(k)=2\pi/k$ becomes 
equal to the ``spatial period'' of the Rabi oscillation in Eq. (\ref{eq11}),
or ``Rabi wavelength'' $\lambda_R(k)=2\pi k/k_c^2$. 
For smaller $k$ the quantum aspect of the atomic motion cannot
be ignored, and $k_-$ becomes purely imaginary,
\beq
k_-\approx ik_c, 
\eeq
whereas $k_+$ remains real. 
As a consequence, $e^{ik_-x}$ evanesces, and 
the contribution of $|\lambda_-\ra$  
beyond  
$k_c^{-1}$ (the penetration length for this component)  vanishes,  
%and
%
%\begin{eqnarray}\label{A20}
%\phi_k^{(1)}(x)&\approx&\frac{k}{\sqrt{2\pi}}
%\left\{\frac{e^{ik_c x}}{k_c}
%+\frac{e^{-k_c x}}{ik_c}\right\}
%\\ \nonumber
%\phi_k^{(2)}(x)&\approx&\frac{-k}{\sqrt{2\pi}}
%\left\{\frac{e^{ik_c x}}{k_c}
%-\frac{e^{-k_c x}}{ik_c}\right\}
%\end{eqnarray}
%
%
\begin{eqnarray}
\phi_k^{(1)}(x)&\approx&\frac{C_+}{\sqrt{2\pi}
}e^{ik_+x}, \;\;\;x>k_c^{-1}  
\label{klkc}
\\ \nonumber
\phi_k^{(2)}(x)&\approx&-\frac{C_+ (\delta+\Omega')}{\sqrt{2\pi}
\Omega}e^{ik_+x},\;\;\;x>k_c^{-1}. 
\end{eqnarray}
This results in a pure, and non-oscillating internal state, as discussed 
later in Section \ref{evdis}.

\section{Temporal  Rabi oscillations:  velocity effects \label{evros}}
In addition to the separation between ``classical'' and ``quantum'' regimes 
at $k_c$, see Eq. (\ref{kc}),  
the classical regime
may be also subdivided into intermediate and high velocities,
depending on whether 
or not the temporal Rabi oscillations are suppressed.
(We shall see later than for spatial oscillations this 
later subdivision does not apply.)   
%We shall in summary distinguish three basic velocity
%regimes for ``high'', ``intermediate'', 
%and ``low energies''.   

\subsection{High velocities: $\la k\ra >k_R$}

Suppose that   
the center of mass motion of the wave packet can be reproduced with
an ensemble of classical 
atoms, as discussed in the previous section, and that  
they   
enter into the laser region in a very short time compared to 
the the Rabi period $T_R\equiv 2\pi/\Omega'$ (sudden entrance).
In that case  they will 
oscillate in phase and the Rabi oscillations will be seen, as 
in the dotted line of Figure \ref{resumen}. 
For Gaussian packets, the transition 
(average) wave number between adiabatic and sudden entrance regimes 
may be identified by equating $T_R$ and a measure of the duration 
of the 
wave packet passage  across the origin,
$t_p\approx 5\Delta_x m/\la k\ra\hbar$, 
where $\Delta_x$ is the wave packet width (square root of variance)
at the peak's passage across $x=0$,  
and the factor 5 is rather arbitrary but is only intended to give an 
estimate, 
\beq
k_R\approx \frac{5 \Delta_x m\Omega'}{2\pi\hbar}. 
\label{esti}
\eeq

\subsection{Intermediate velocities: $k_c<\la k\ra <k_R$}
In the intuitive language suggested by the 
classical approximation for the center of mass motion, the ``mechanism''
that explains the suppression of temporal Rabi oscillations at
intermediate velocities is the averaging of the Rabi oscillations of
the individual atoms forming the ensemble when the entrance time 
for the whole ensemble is
long compared to the  Rabi period. Since each atom enters at a different 
instant in the laser region, the phase of the oscillation will be 
also different. This is the case shown in Fig. \ref{densidades1}, where 
the peaks of the densities of the ground and excited state alternate
with a spatial period given by $\hbar \la k\ra T_R/m$.

\begin{figure}
{\includegraphics[width=3.35in]{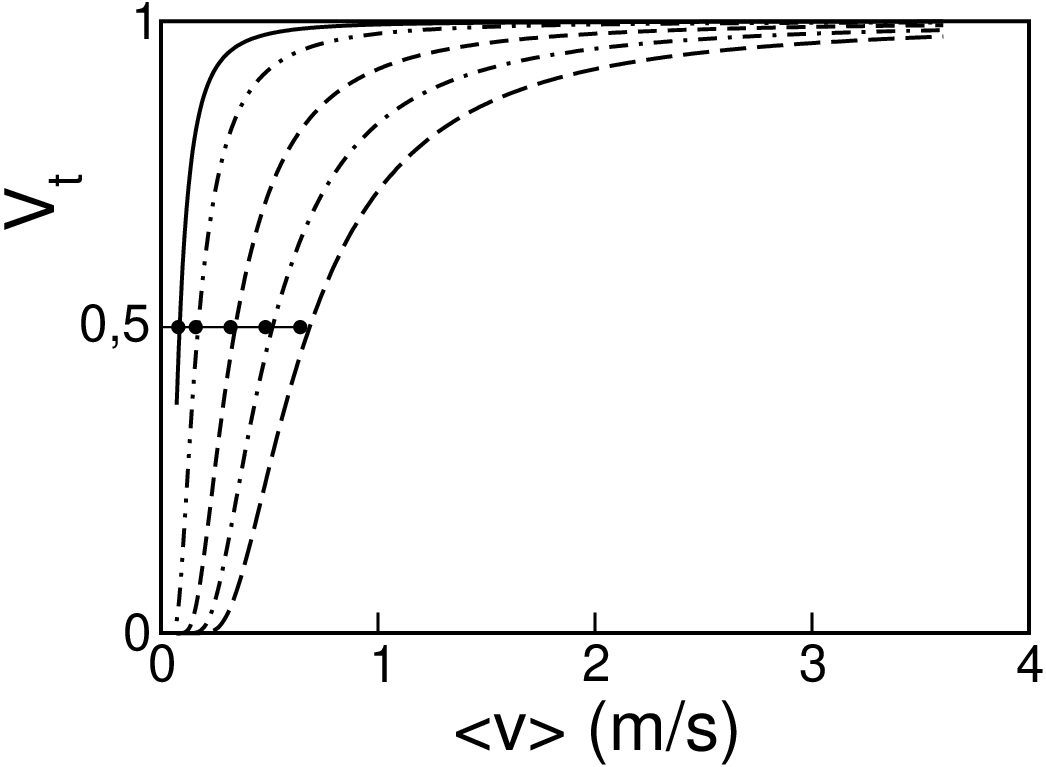}}
\caption{Visibilities, Eq. (\ref{vis})
(the quantum and semi-classical results 
are indistinguishable in the scale of the figure)
for (from left to right)  
$\Omega=0.413,\, 0.827,\,
1.654,\,2.480,$ and $3.307$  $\times 10^6$ s$^{-1}$.  
In all cases the initial Gaussian has $\Delta_x=0.2438\,\mu$m,
$\la x\ra=-1.32\,\mu$m, and $\gamma=\delta=0$. 
Also drawn are the corresponding values of the transition velocities
$v_R= k_R\hbar/m$ with circles.}
\label{omega}
\end{figure}

Figure \ref{omega} shows the ``visibility'' $V_t$
obtained from maxima (max) and minima (min) of the probability
of the excited state,
$P_2(t)$, 
corresponding to large times beyond the initial transient,
\beq
\label{vis}
V_t=\frac{P_2(max)-P_2(min)}{P_2(max)+P_2(min)},
\eeq
versus the incident average velocity for five different values of $\Omega$ 
and a fixed wave packet width, 
as well as the estimate given by Eq. (\ref{esti}); 
again $\delta=0$. The  visibilities
calculated exactly or with the classical approximation
are indistinguishable.   
%Clearly this transition criterion will hold as long
%as the classical approximation 
%is valid on both sides of the transition, namely, when $k_R>k_c$,
%which we have assumed all along. The case $k_c>k_R$ is in principle 
%possible but quite unlikely to find in practice.  
%We shall always assume that the 
%opposite case, $k_R>k_c$, holds. 

%Note that the above semiclassical ``explanation'' hinges on a premise
%that is manifestly false, 
%namely, the translational part of the state is not a classical
%ensemble 
%but quantum and coherent for all incident energies. 
%%For some applications, as we have seen, 
%%ensemble 
%%but to a quantum and coherent state for all incident energies. 
%Nevertheless, as we have seen, 
%the formal substitution by a classical mixture is harmless 
%at sufficiently high energies. 
%(By the way, true quantum statistical mixtures,
%more likely to be found in an actual experiment 
%than the pure state considered so far, would also lead to a translational 
%ROS effect.) 
%A difference with a truly 
%classical motion is that 
%the subensemble of semiclassical atoms associated with a single phase space 
%point, i.e., a given classical trajectory, would oscillate according to Rabi's
%frequency, with 
%an observable consequence in the 
%fluorescence signal. Of course no such a subensemble may be formed  
%in the 
%quantum case. 

%
%see Fig \ref{sw1}.
%
%\begin{figure}
%\label{sw1}
%{\includegraphics[width=3.35in]{sw1n.eps}}
%\caption{Densities of ground and excited state versus $x$ for $p=44$ and
%$\Omega=800$.}
%\end{figure}
%

\subsection{Low velocity: $\la k\ra<k_c$}
%When the classical approach fails 
%the Rabi oscillation is still suppressed, but
%the mechanism is 
%now rather different.

%
Below $k_c$ there are no Rabi oscillations either but, 
according to the discussion of section \ref{sa}, the
reason is not an
effective ``averaging'' (the semiclassical approximation is not valid), 
but the fact that the atomic internal state formed by the laser does not
oscillate at all.

\section{Internal states: Effects of the velocity  and detuning \label{evdis}}
The motion of the atom from the laser-free region to the laser-illuminated
region has very different effects on the internal state depending on
the incident velocity and the detuning.  
We shall discuss now the effect of the various velocity regimes in 
the degree of mixing of the normalized, internal,
reduced density operator
for the atoms in the laser region,
which is defined by its matrix elements 
\beq
\label{rol}
\rho^L_{ij}(t)=\frac{\into dx\,[\psi^{(i)}(x,t)]^*
\psi^{(j)}(x,t)}{
\into dx\,|\psi^{(1)}(x,t)|^2 + |\psi^{(2)}(x,t)|^2}, 
\eeq
distinguishing between the cases of zero and non-zero detuning.

\subsection{Zero detuning}
%
%We shall now concentrate on the case $\delta=0$, 
%and examine the internal states created in 
%regimes of low, intermediate and high wave numbers.

({\it i}) For $  \la k \ra <k_c$ the populations of both ground and
excited states are equal and the overlap in space of the respective wave
functions is maximum.  After a transient time, necessary for
Eq. (\ref{klkc}) to apply,  
$\rho^L$ is finally given by the pure state $2^{-1/2} (|1\ra-|2\ra)$, 
\beq
\rho^L=\left({1/2\atop{-1/2}}{{-1/2}\atop 1/2}\right),
\eeq
which is nothing but $|\lambda_+\ra\la\lambda_+|$, now  normalized.

\begin{figure}
{\includegraphics[width=3.3in]{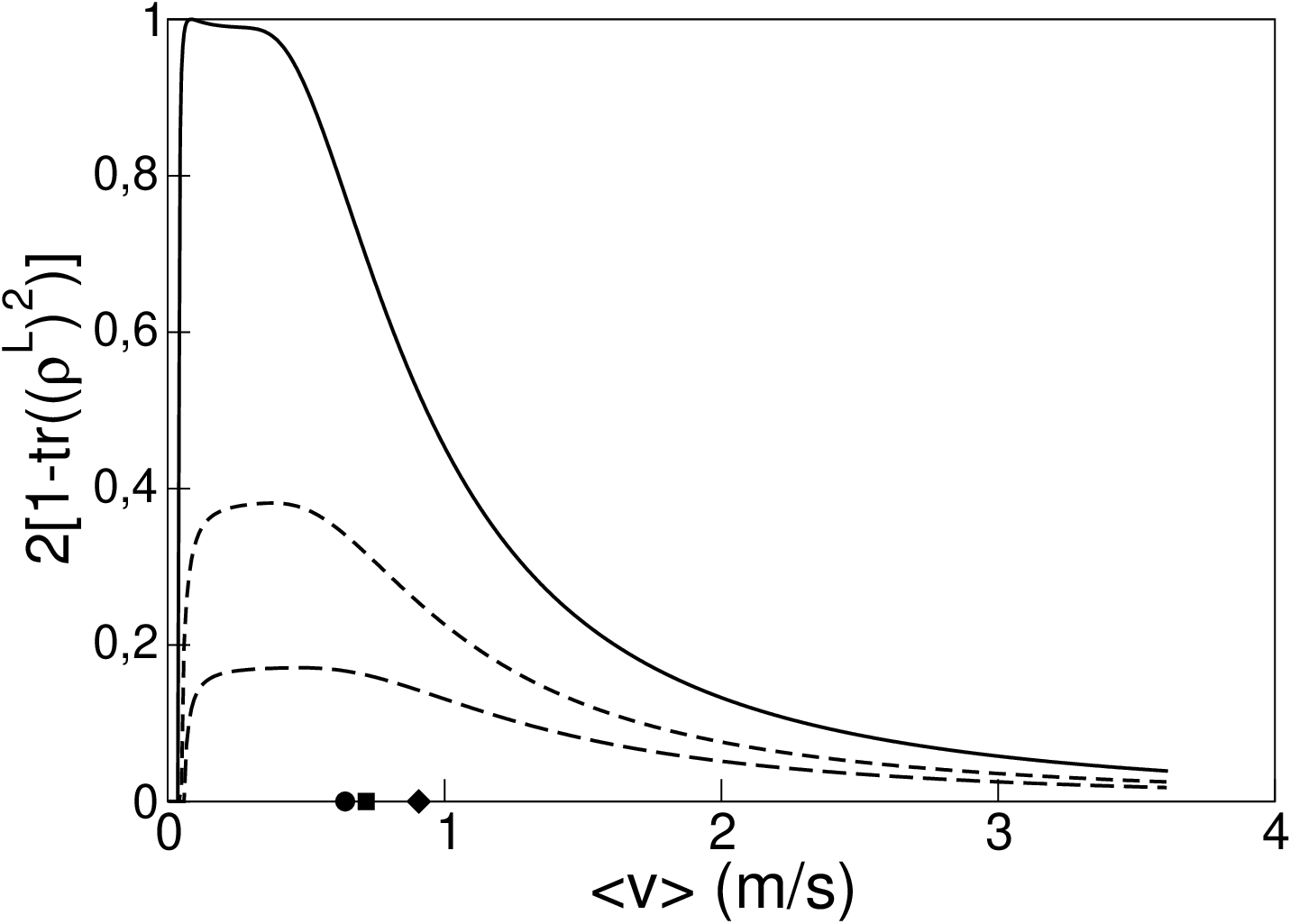}}
\caption{Degree of mixing versus $\la v\ra$  
for $\gamma=0$,
$\Omega=3.3 \times 10^6$ s$^{-1}$, $\la x\ra=-1.34\,\mu$m, and
$\Delta_x=0.2436\,\mu$m.     
$\delta=0$ (solid line), $-\Omega/2$ (short dashed line) and $-\Omega$
(long dashed line). 
Also marked are the corresponding values 
of $\hbar k_R/m$ (circle, square and diamond respectively).\vspace*{2.cm}\\   
}
\label{mezcla3}
\end{figure}

({\it ii}) For $k_c<k<k_R$ excited and ground components do not overlap,
see Fig. \ref{densidades1},
so $\rho^L$ tends to a diagonal matrix. 
The two populations are still equal (there
is no global Rabi oscillation because of the semiclassical averaging effect)
and $\rho^L$ is 1/2 times the unit matrix, which amounts to the
maximum degree of mixing (incoherence) allowed for the reduced internal state.
The sharp transition from a pure state to a maximally
mixed state around 
$\la k\ra =k_c$ may be seen in Figure \ref{mezcla3}.

\begin{figure}
%$^{}$\vspace*{1.cm}\\ 
{\includegraphics[width=3.35in]{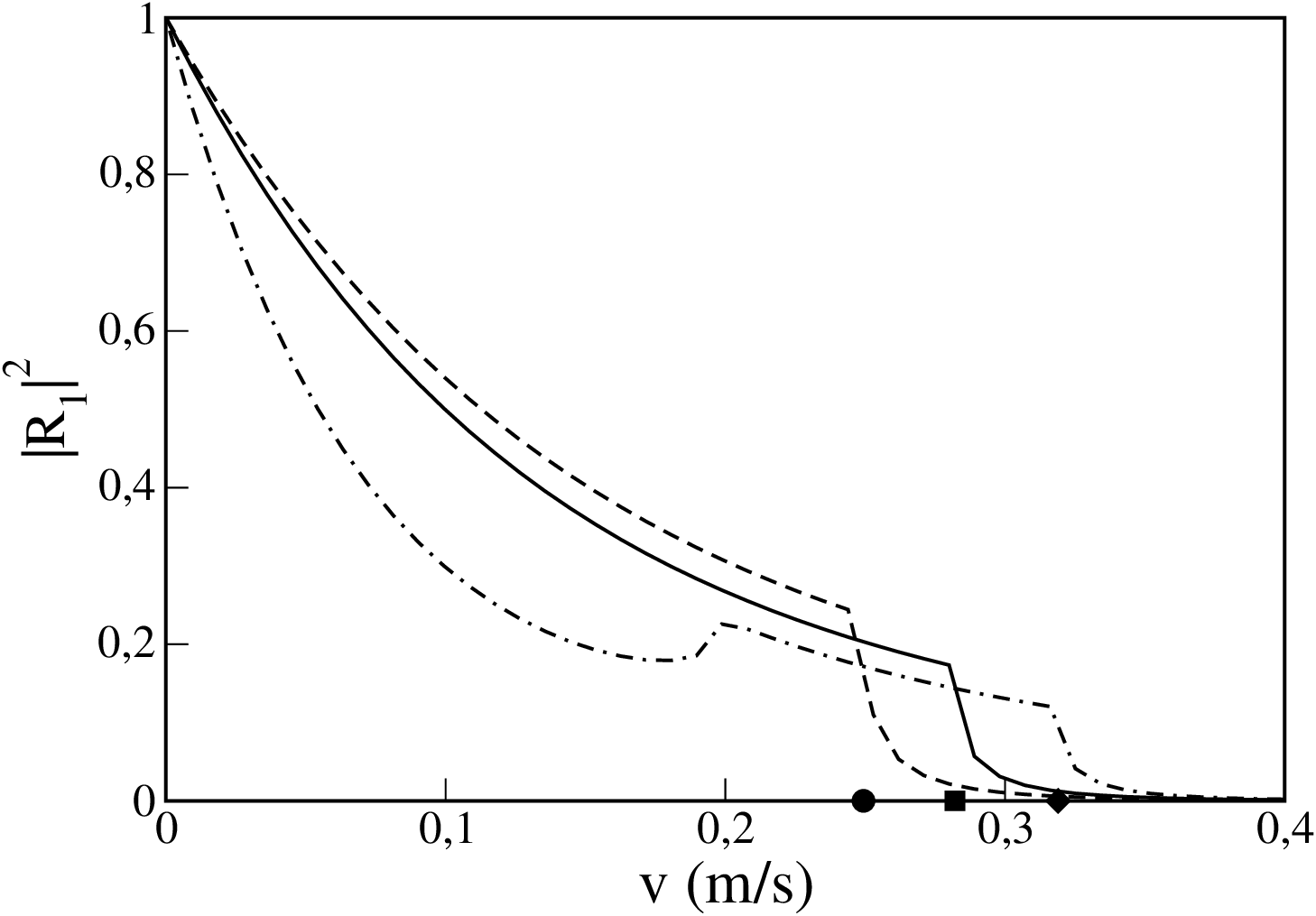}}
\caption{Reflection probability to the ground state versus $v$
for $\Omega=166.5 \times 10^6$ s$^{-1}$; $\gamma=0$;
$\delta=$ $-\Omega/4$ (dotted-dashed line), 
$0$ (solid line), and $\Omega/4$ (dashed line). 
$\hbar k_c/m$ is indicated with a circle, square and diamond respectively. 
}  
\label{r31a}
\end{figure}

\begin{figure}
%$^{}$\vspace*{1.cm}\\ 
{\includegraphics[width=3.35in]{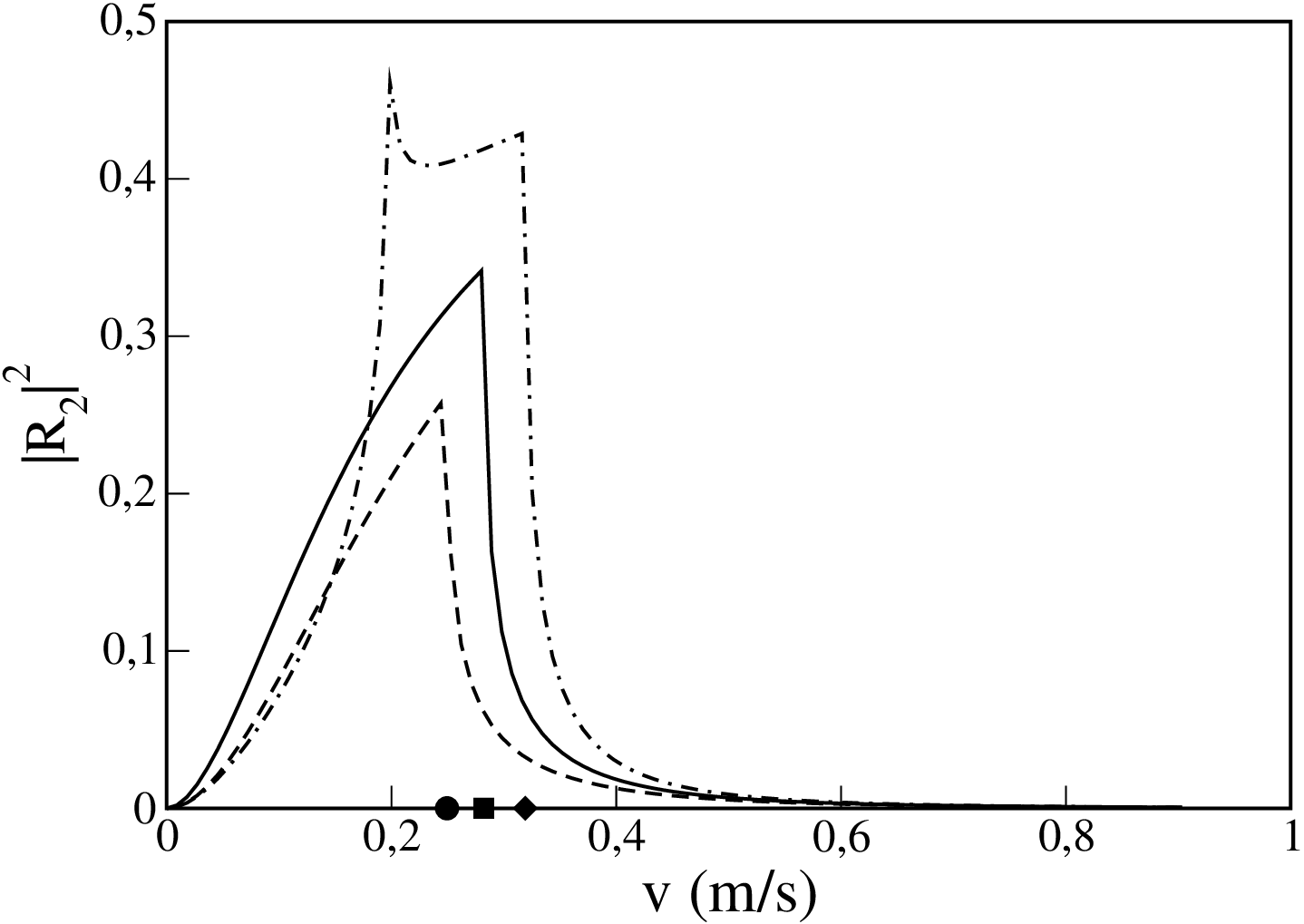}}
\caption{Reflection probability to the excited state versus $v$
for $\Omega=166.5 \times 10^6$ s$^{-1}$; $\gamma=0$; 
$\delta=$ $\Omega/4$ (dashed line), 
$0$ (solid line), and $-\Omega/4$ (dotted-dashed line). 
$\hbar k_c/m$ is indicated with a circle, square and diamond respectively.
The peak close to $0.2$ is at the point where $q$ passes
from imaginary to real.   
}   
\label{r32a}
\end{figure}

There is also quite  a dramatic effect of $k_c$ on the reflection
probabilities, which tend to vanish for higher values of $k$, 
see Figures \ref{r31a} and \ref{r32a}
for $|R_1|^2$ and $|R_2|^2$ respectively. 


({\it iii}) Finally, for $k\gg k_R$ there is no classical averaging effect, 
and the two populations oscillate periodically and coherently
in time between 0 and 1. The reduced internal state in the laser region 
becomes pure again, as
Figure \ref{mezcla3} shows. Note that the transition from intermediate to 
high kinetic energies at $k_R$ is rather smooth, in particular in comparison 
with the sharp $k_c$-transition.  

If the
initial atom is prepared in the
excited state instead of the ground state,
different scattering eigenfunctions 
have to be used. The calculation  
follows similar lines
and provides different expressions for reflection amplitudes and coefficients 
$C_\pm$. However, the eigenstates $|\lambda_\pm\ra$ remain the same, 
and it is still true that below $k=k_c$ only $k_+$ is real, so that   
the same pure state is obtained at low kinetic energy.  
This means that the low-energy 
motion of the atom into a laser illuminated region 
provides a very simple physical  
mechanism to project any initial internal state, pure or mixed, 
onto the 
pure state $2^{-1/2} (|1\ra-|2\ra)$ if $\delta=0$. The laser
illuminated region may in fact be
moved and the atom be at rest with the same effect. 

\subsection{Non-zero detuning}
Detuning may be used as a control knob to obtain states with
a smaller fraction of excited state.    
The (unnormalized) pure state obtained at low kinetic energy in the laser
region is given 
for arbitrary $\delta$ by 
\beq
{\bf \Psi}=
{1\choose-\frac{\delta+\Omega'}{\Omega}}.
\eeq
As seen in Figure \ref{mezcla3},  
for the intermediate regime $k_c<\la k\ra <k_R$, 
the classical averaging is not so effective for 
non zero detuning, so that the state is less mixed.    
Note also that negative detuning leads to less reflection at very low
kinetic energies than positive 
detuning, see Figs. \ref{r31a} and 
\ref{r32a}, and the existence of an additional critical
point associated 
with the transition between imaginary and real values of $q$.  

\section{Suppression of spatial Rabi oscillations \label{rospd}}

The quantum suppression of Rabi oscillations below $k_c$ may also
be seen in several space dependent quantities.
They are, however, not sensitive at all to the
semiclassical mechanism: these quantities oscillate both below and
above $k_R$ defined in Eq (\ref{esti}).  
 
\begin{figure}
%$^{}$\vspace*{2.cm}\\ 
{\includegraphics[width=3.35in]{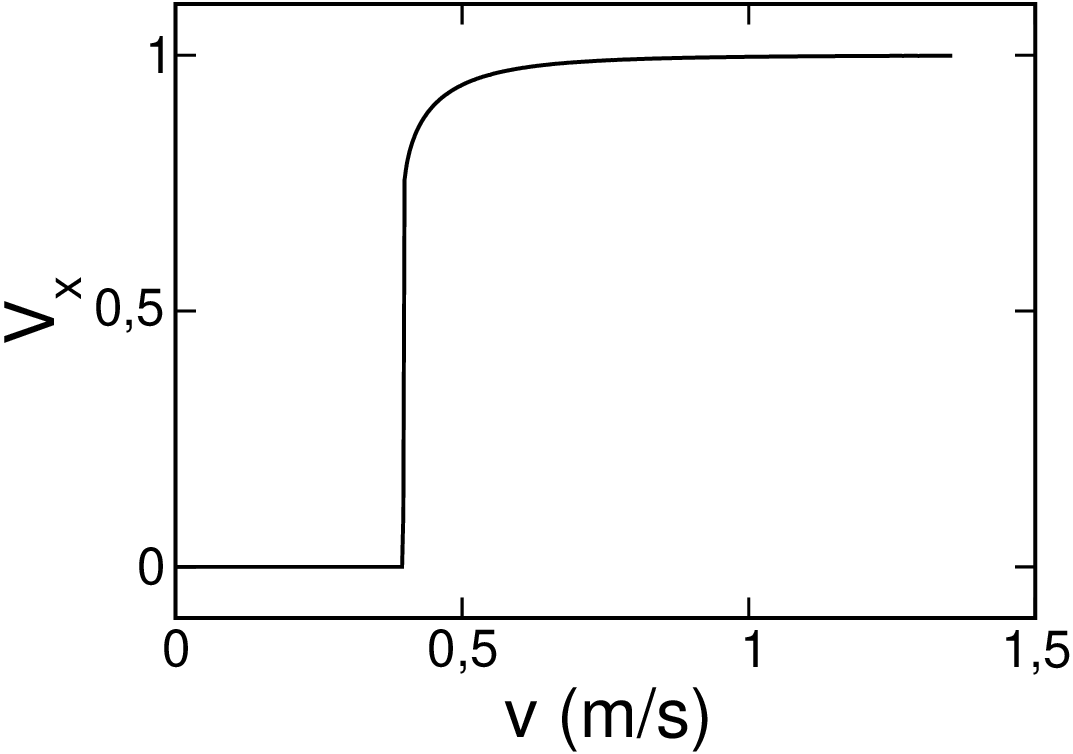}}
\caption{Spatial visibility, Eq. (\ref{visix}), versus $v$ for
$\Omega=3.307 \times 10^6$ s$^{-1}$, $\gamma=0$,
and $\delta=0$.\vspace*{.0cm}\\ 
}  
\label{sw3}
\end{figure}

In Figure \ref{sw3} we show 
\beq
\label{visix}
V_x=\frac{|\phi_k^{(2)}(x)|^2(max)-|\phi_k^{(2)}(x)|^2(min)}{
|\phi_k^{(2)}(x)|^2(max)+|\phi_k^{(2)}(x)|^2(min)}
\eeq
for maxima and minima with respect to $x$ evaluated 
beyond the penetration length of the $\lambda_-$-component.
One may see a rather abrupt 
jump from 0 to 1 at $k=k_c$.

A similar jump may be observed for $I(x)$ of
Eq. (\ref{taux}). For  wave packets with components below $k_c$,
so that $k_+\approx k_c,\, k_-\approx ik_c$, 
and for $x$ beyond the transient, small-$x$ region,
the $t$ integral can be done to obtain $I(x)\approx \la v\ra/v_c^2$.  
On the other hand, for $k$ above $k_c$, $I(x)$ becomes  
\beq
I(x)\approx \frac{m}{\hbar}\into\,\frac{dk}{k}\sin^2
\left(\frac{\Omega x}{2v}\right)|\la k|\psi^{(1)}(0)\ra|^2.
\eeq
These two limiting cases are depicted in Figure \ref{px015}.

%
%
%
%
%\begin{figure}
%{\includegraphics[width=2.7in]{mezcla1.eps}}
%\caption{Degree of mixing versus $\la v\ra$  
%for 
%$\Omega=3.3$MHz, $\la x\ra=-6.34\mu$m, and $\Delta x=0.2436$.     
%$\delta=0$ (solid line), $\Omega/2$ (short dashed line) and $\Omega$
%(long dashed line). 
%Also marked are the corresponding values 
%of $k_R$.   
%}
%\label{mezcla1}
%\end{figure}

\section{Inclusion of Damping \label{sd}}
In this section we shall see that the suppression of Rabi oscillations  
can be detected by means of     
the fluorescence signal of   
the spontaneously emitted photons.
Two cases will be considered: a detection measurement of the 
first fluorescence photon,  
which is described by the conditional Hamiltonian, 
and a measurement where all successive photons are also taken into account, 
which requires a master equation (optical Bloch equations). 
In both cases the fluorescence signal is proportional 
to the population of the excited state $P_2$ which can be  calculated
by means of $H_c$ or by the Bloch equations, respectively.

%Figure 
\begin{figure}
{\includegraphics[width=3.3in]{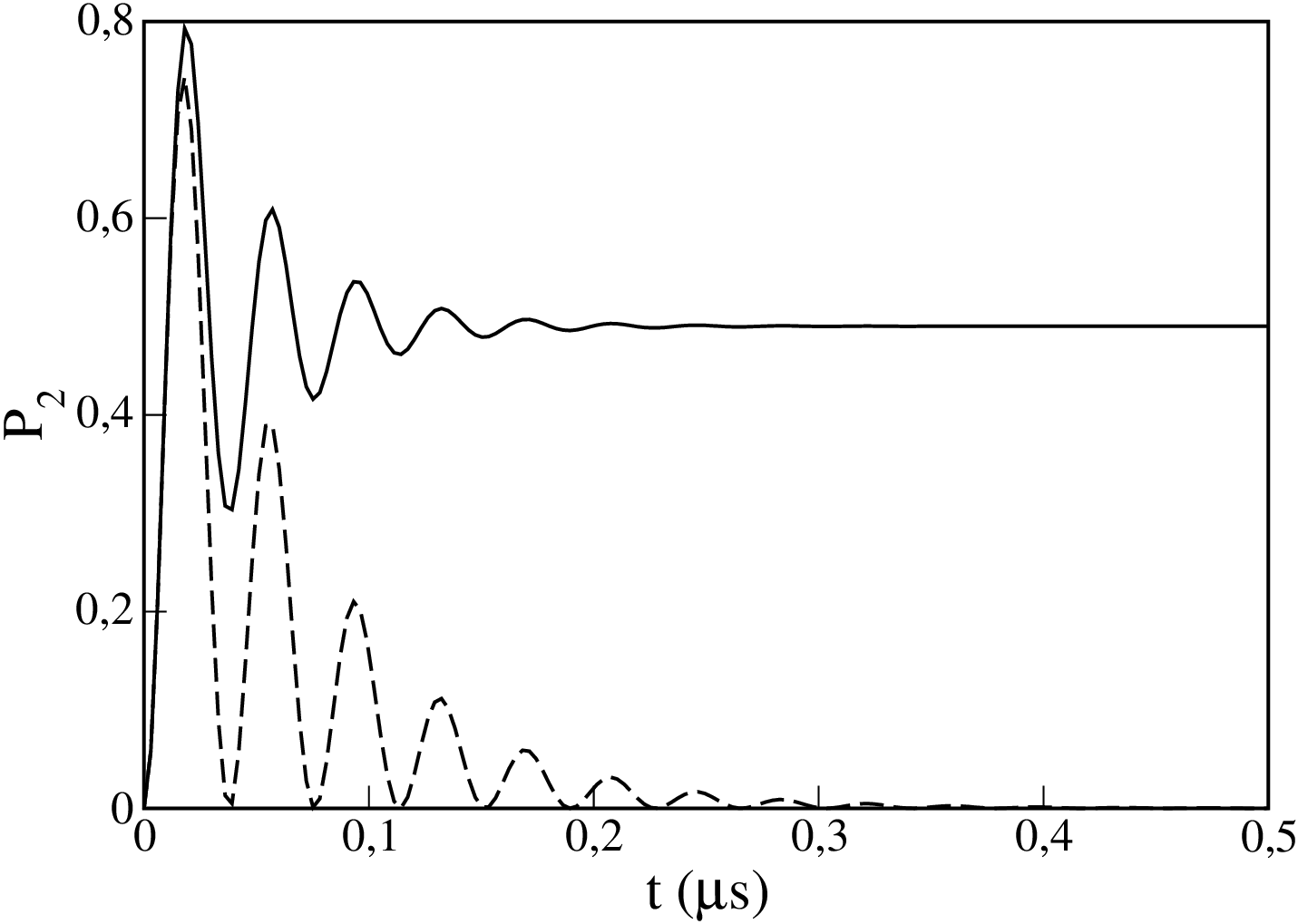}}
\caption{Fading (not suppression!) of Rabi oscillations for an atom at
  rest: $P_2$ versus $t$   for  $\gamma=33.3\times 10^6$ s$^{-1}$,
  $\Omega=5\gamma$,      $\delta=0$.
The solid line corresponds to the master equation and the dashed one 
to the conditional Hamiltonian (``first photon'').  
}
\label{bloch3}
\end{figure}

The suppression of Rabi oscillations discussed here should be
distinguished from their fading away due to damping and  approach to
a stationary internal state. The latter is quite well-known for atoms
at rest \cite{WM} (cf. Figure \ref{bloch3}).
It is possible to operationally distinguish  fading and suppression 
by a combination of $\Omega$ and $\gamma$ that, for the atom at rest,   
allows to observe several oscillations 
before reaching the asymptotic stationary population.
This corresponds to 
strong driving 
conditions, $\Omega\gg \gamma$.

\subsection{Rabi oscillations in the first-photon distribution}
For the one-dimensional case the conditional Hamiltonian $H_c$ can be
written as
\beq
H_c= \frac{\hat{p}^2}{2m}+\frac{\hbar}{2}
\left({0\atop \Omega\Theta(\hat{x})e^{ik_Ly}}
{\Omega\Theta(\hat{x})e^{-ik_Ly}
\atop -2\delta-i\gamma} \right).
\label{hamga}
\eeq
The generalization of Eqs. (\ref{A14}-\ref{d}) for $\gamma\ne 0$
is provided in Appendix \ref{decay}.

\begin{figure}
{\includegraphics[width=3.3in]{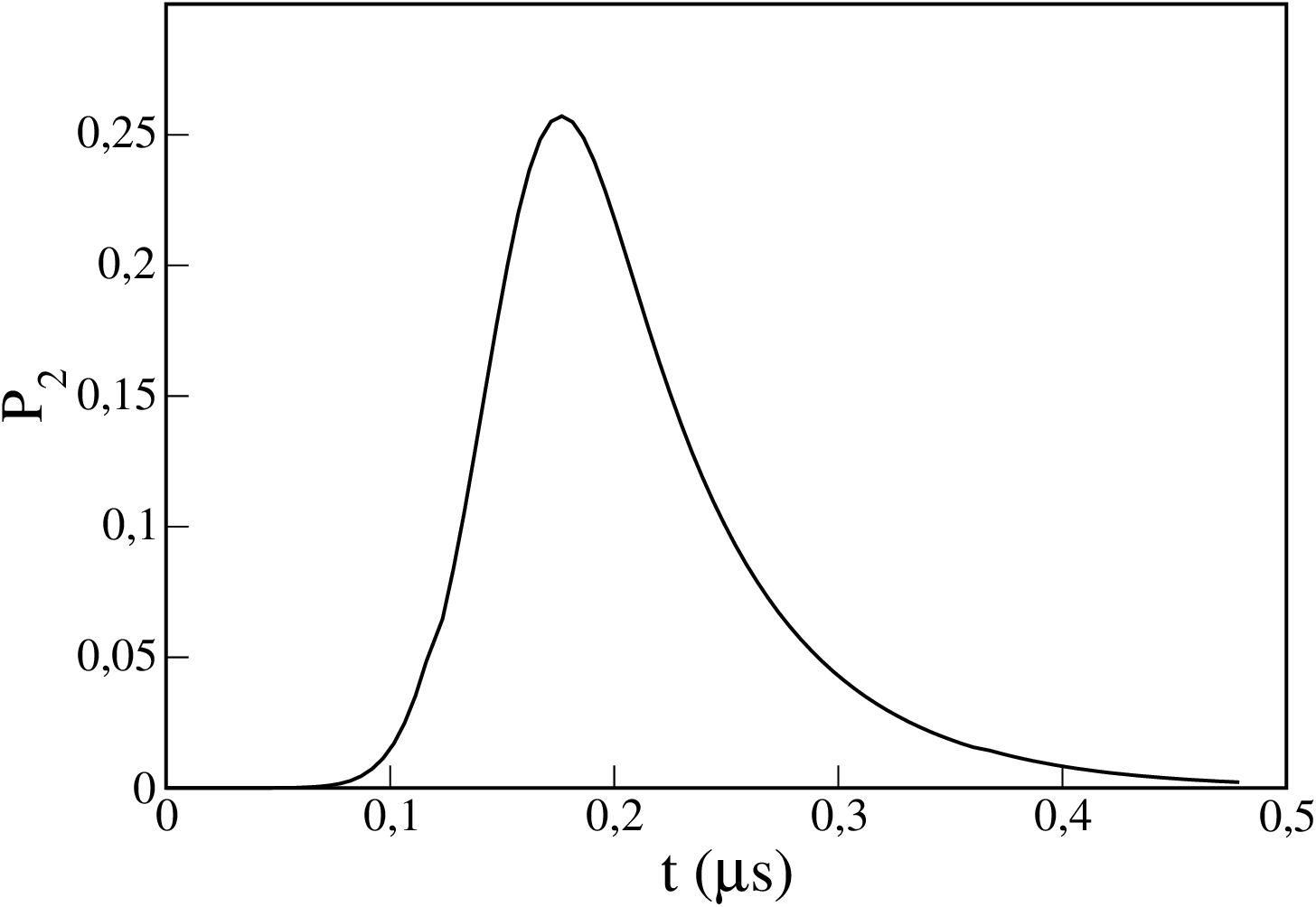}}
\caption{$P_2$ versus $t$  
for  $\gamma=33.3\times 10^6$ s$^{-1}$,
$\Omega=5\gamma$ and      
$\delta=0$. The parameters of the initial Gaussian are 
$\la v\ra=9,03$ m/s, $\Delta_x=0.2436$ $\mu$m, and $\la x\ra=-1.322$ $\mu$m.
}
\label{gamma2}
\end{figure}

For high velocities the entrance of the 
wave packet in the laser region is sudden, so that $P_2$ versus time
takes 
the same form as for the atom at rest from the entrance time on.  
At lower velocities the oscillation pattern 
disappears; 
$k_R$ still marks the transition, which is now 
smoother than when $\gamma=0$.   
Figure \ref{gamma2} shows $P_2$
for a velocity below $k_R$. The Rabi oscillation suppression
is evident, compare 
with the dashed line of Figure \ref{bloch3}. 

\begin{figure}
{\includegraphics[width=3.3in]{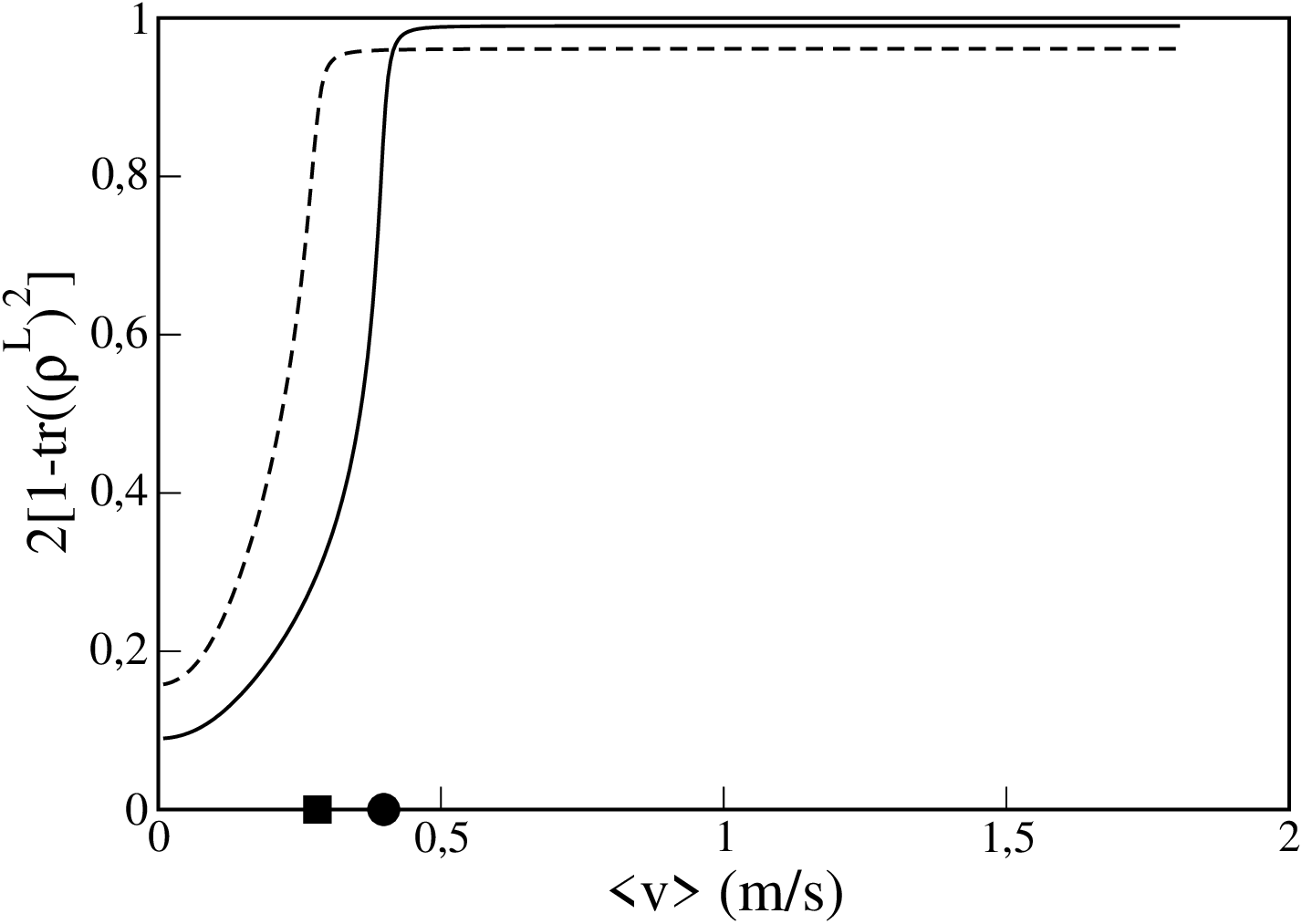}}
\caption{Degree of mixing of the internal state in the laser region 
versus $\la v\ra$ for $\Delta_x$= 0.2436 $\mu$m, 
$\gamma=33.3\times 10^6$ s$^{-1}$, 
$\Omega=10\gamma$ (solid line), and 
$\Omega=5\gamma$ (dashed line).
The corresponding values of $\hbar k_c/m$ are also 
shown with a circle and a square respectively.} 
\label{swanalitic1}
\end{figure}

The transition at $k_c$ is illustrated in Fig. \ref{swanalitic1},
which shows the
degree of mixing versus the average
velocity \cite{foot2}. Note the smoothing effect 
of a non-zero $\gamma$.

An observation of the suppression of Rabi oscillations using the
first photon could be 
achieved for a ``Lambda'' configuration of three
atomic levels such that the laser  
couples two of them, while a third one acts as a sink for the 
excited state \cite{EFYS02} (only one photon may be emitted per atom).
The ideal experiment would require 
a preparation of the atoms, sent one by one, according to a single 
wave function, and   
a detector capable of responding to a single photon. 
The experiment would be repeated many times until a profile similar to 
the one in Fig. \ref{gamma2}
is obtained for the density of photon detection times.

%Figure 

%Figure 

\subsection{Temporal Rabi oscillations in the photon intensity}

For a simple two-level atomic configuration, without a sink level for the 
excited state, the atom may emit many photons. 
In this case   
recoil effects
may become important, 
especially at low incident energies, and lead to 
spatial broadening.   
The master equation 
that describes the density operator $\rho^{3D}$ of the atomic system
in three spatial dimensions  is given, following \cite{Hegerfeldt93}, by
\beq
\label{master}
\dot{\rho}^{3D}= 
-\frac{i}{\hbar}\{H_c\rho^{3D}-\rho^{3D} H_c^\dagger\}
+{\cal R}(\rho^{3D}), 
\eeq
where $H_c$ is the conditional 3D Hamiltonian, Eq. (\ref{2.4}),  
and the ``reset'' or ``jump'' term ${\cal R}(\rho^{3D})$
takes into account the atomic
recoil along the directions weighted by the dipole emission 
distribution. In momentum representation,   
\beqa
\la {\bf p}|{\cal R}(\rho^{3D})|{\bf{p}}'\ra&=&
\gamma |1\ra\la 1|
\\
&\times&
\int d{\bm{\kappa}}
P(\bm{\kappa})
\la {\bf{p}}+\hbar k_L{\bm{\kappa}}|\rho^{3D}_{22}
|{\bf p}'+\hbar k_L{\bm{\kappa}}\ra,
\nonumber
\eeqa
where the integral is over all possible photon
directions represented by the unit 
vectors $\bm{\kappa}={\bf k}/|{\bf k}|$, and 
\beq
P({\bm{\kappa}})=\frac{3}{8\pi}\left(1-\frac{{\bm{\kappa}}
\cdot {\bf d}}{|{\bf d}|^2}
\right)=\frac{3}{8\pi}\sin^2\theta,
\label{dis}
\eeq
$\theta$ being the angle between ${\bm{\kappa}}$
and the dipole moment ${\bf d}$, which is taken
along the $z$ direction.   

Even though the fluorescence photons and the associated atom recoil may
go in any direction,  
one can take the trace over momentum components $p_y$ and $p_z$
in the reset term, 
and reduce the integral to a one dimensional one, see e.g. \cite{Hensinger01},
to obtain 
for the initial atomic-motion direction $x$ the expression 
\beqa
{\cal R}_x(\rho)&\equiv& {\rm Tr}_{y,z} {\cal R}(\rho^{3D})=|1\ra\la 1|
\gamma\frac{3}{8}
\\
&\times&
\int_{-1}^{1}du\, (1+u^2)
\la p_x+\hbar k_L u|\rho_{22}|p'_x+\hbar k_L u\ra,
\nonumber
\eeqa
where the reduced atomic density operator for the $x$-direction is   
\beq \label{reduced}
\rho_{ij}\equiv {\rm Tr}_{y,z} \la i|\rho^{3D}|j\ra,\;\;i=1,2,\;\;j=1,2. 
\eeq
Note that it contains both internal and
center of mass information of the atom, so it is different from 
$\rho^L$ in Eq. (\ref{rol}). 
Even though, taking the trace over $y$ and $z$ 
in the other terms of Eq. (\ref{master}) does not lead to a closed
equation for the reduced one-dimensional
density matrix of Eq. (\ref{reduced}), 
a one-dimensional approximation is valid,
\beq
\label{master1}
\dot{\rho}= 
-\frac{i}{\hbar}\{H_c\rho-\rho H_c^\dagger\}
+{\cal R}_x(\rho), 
\eeq
with $H_c$ given by Eq. (\ref{hamga}),
provided that the 
effect of the two broadening mechanisms in the $y$ direction,     
the standard quantum mechanical wave packet spreading
and the atomic recoil, remain small.   
The time scales of both effects are estimated in 
Appendix \ref{limits}.   

\begin{figure}
{\includegraphics[width=3.3in]{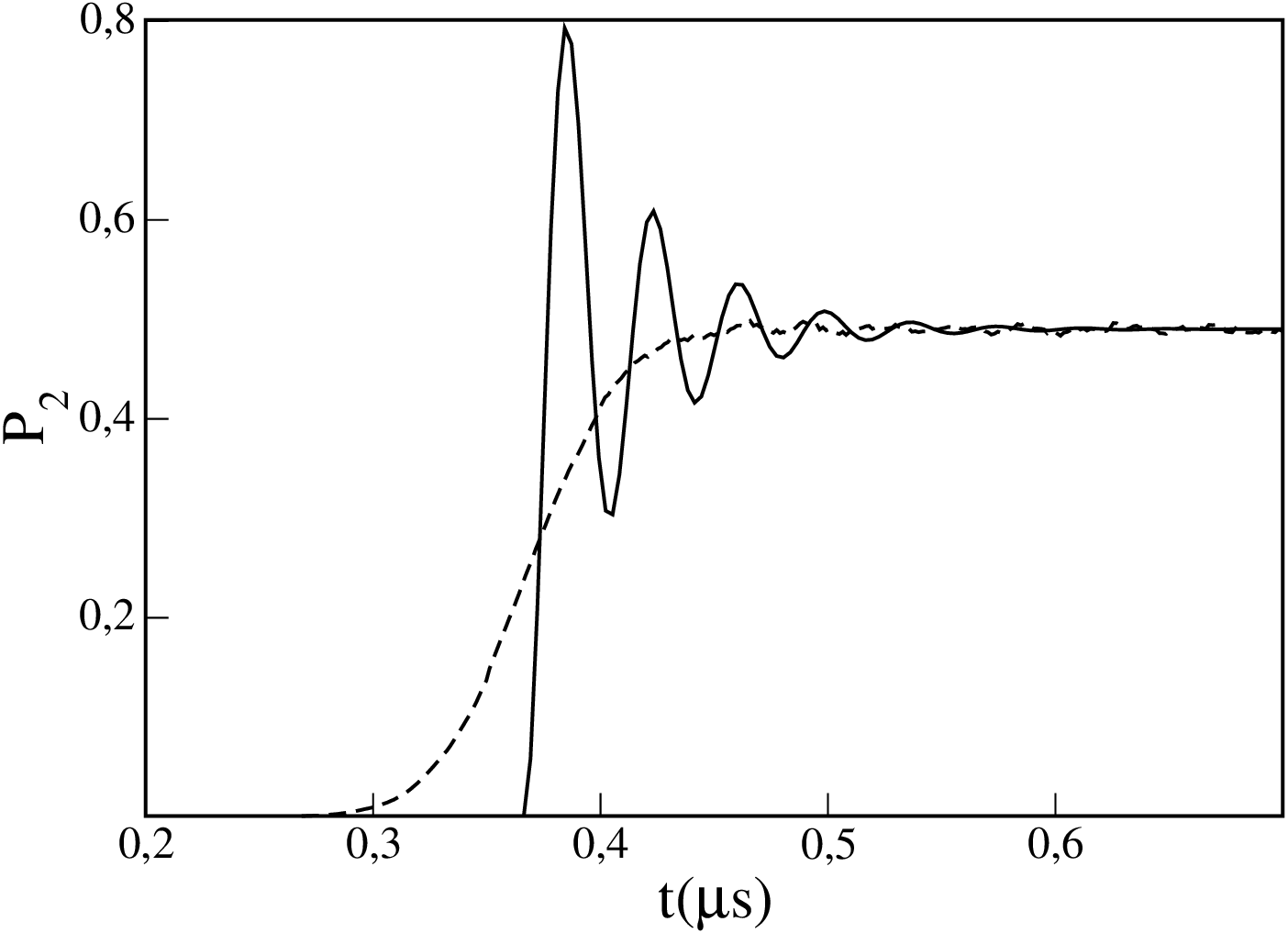}}
\caption{$P_2$ versus time  
for $\gamma=33.3\times 10^6$ s$^{-1}$, $\Omega=5\gamma$, 
$\Delta_x$= 0.12 $\mu$m, $\la x\ra=-1.322\mu$m, $\la v\ra=3.61$ m/s;
average for 10000 trajectories (dashed line), compared to $P_2$ for a very fast 
packet (solid line). The later curve corresponds to the 
solid line of Figure \ref{bloch3},
displaced so that the starting time 
coincides with the entrance of the center of the wave packet in the laser
region.} 
\label{judas}
\end{figure}

The fluorescence signal is proportional to $P_2$, which is now the 
total population 
of excited atoms, regardless of whether or not they have emitted.  
The one dimensional master equation 
may be solved in principle directly, or, as done here, using the quantum 
jump technique, i.e., averaging over many ``trajectories'' \cite{Hegerfeldt91}.
For each of them 
the photon detections occur at random instants. 
Figure \ref{judas} shows $P_2$  averaged over 10000
trajectories, compared with the solid line of Figure \ref{bloch3}, properly 
shifted, which gives the time evolution of $P_2$ 
for a sudden entrance of a very fast wave packet.  
The suppression of Rabi oscillations is evident. 

The ideal fluorescence measurement would be performed 
by sending one atom at a time, as described before. In a less demanding 
experiment, one could prepare a cloud of many atoms of sufficiently 
low density, 
%so that the free-motion approximation is still valid, 
and send it towards the laser, measuring the fluorescence 
signal versus time. The cloud ensemble would lead to 
suppression of Rabi oscillations at sufficiently low velocities. 
In this case the incident translational state would not be pure 
but a true mixture.     
%Similar settings may be conceived for the case when one single atom 
%may emit many photons in repeated pumping and emission cycles.

The transition at $k_c$ would be most easily noticed by the reflection
dominant at lower kinetic energies. For the Cs transition we are considering 
in the numerical examples $v_c\equiv\hbar k_c/m=0.28$ m/s,
which is well above the recoil
velocity limit $\hbar k_L/m=0.35$ cm/s. The possibility to control 
the pure state formed in the laser region will depend 
on the ability 
to focus the atomic beam along the $y$- direction in the scale of the laser 
wavelength, due to  
the $y$-dependence of the  
pure state obtained, see Eq. (\ref{decayf}).   

%Figure 

\subsection{Spatial Rabi oscillations \label{spatial}}
With damping included, $P_2(x,t)$ becomes  $\la
x|\rho_{22}(t)|x\ra$ so that $I(x)$ in Eq. (\ref{taux}) becomes
\beq
I(x)= \int_{-\infty}^\infty dt\,\la x|\rho_{22}(t)|x\ra ~,
\label{pofx2}
\eeq
and $\gamma I(x)$ is the mean number of photons per atom per unit length.

In a similar way, one may consider the probability density, denoted by
$\gamma I_0(x)$, for the atomic position when  the first 
photon is emitted. 
%This is obtained by considering the joint
%probability density that the first photon is emitted at time $t$
%and the atom
%is at position $x$.
Similar to Eq. (\ref{taux}),  $I_0(x)$
is given by
\beq
I_0(x)= \int_{-\infty}^\infty dt\,|\psi^{(2)}(x,t)|^2,
\label{pofx1}
\eeq
where the time development is given by the conditional Hamiltonian in
Eq. (\ref{hamga}), and $\psi^{(2)}(x,t)$ denotes the excited-state
component. Integrating Eq. (\ref{pofx1}) over $x$ gives the fraction
of atoms which emit a photon, so that $\int dx\, I_0(x)$ tends to
one for $k\gg k_c$. 

\begin{figure}
{\includegraphics[width=3.3in]{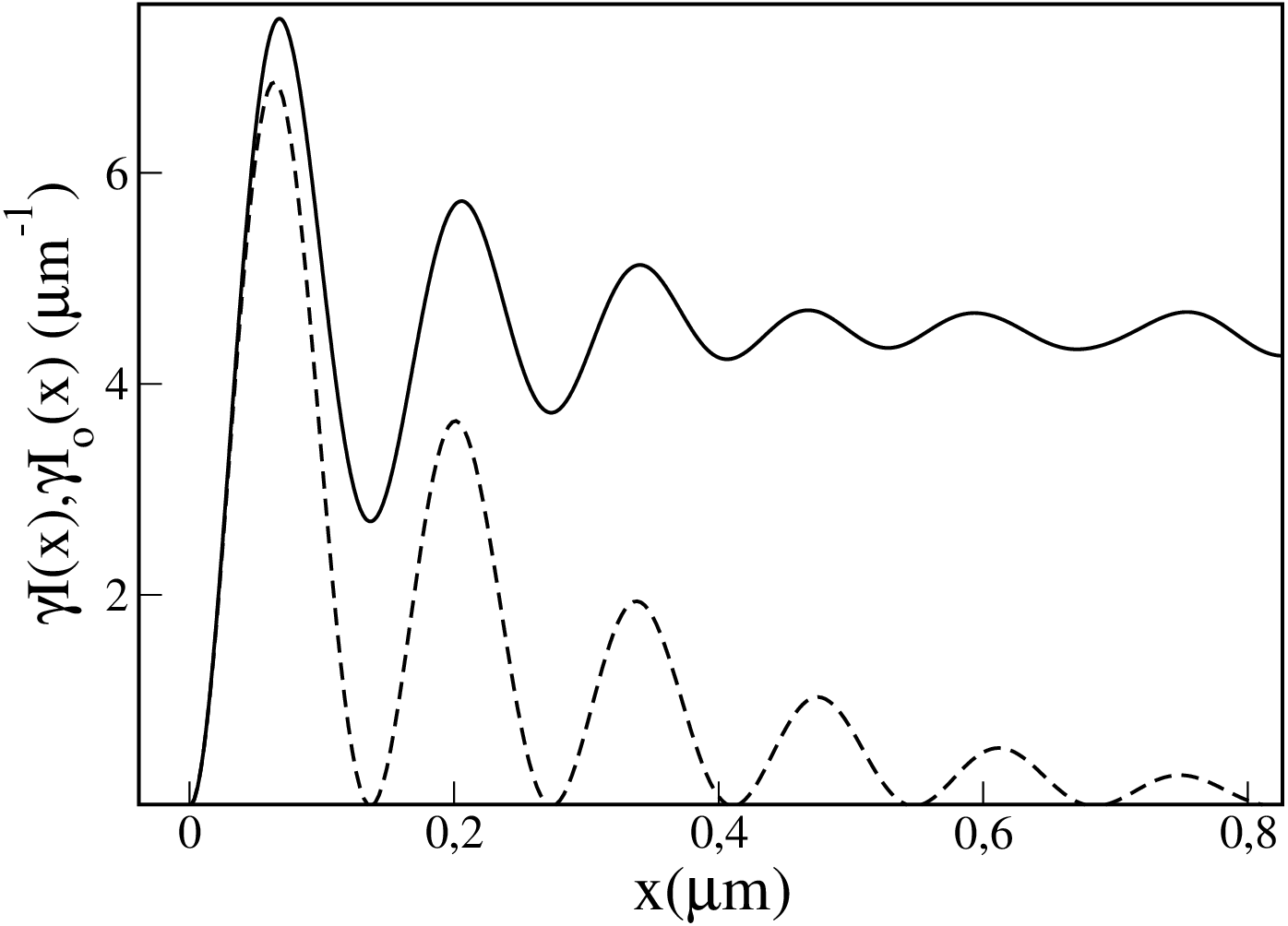}}
\caption{$\gamma I(x)$ (computed with 160 trajectories, solid line)
and $\gamma I_0$ (dashed line) 
versus $x$  
for $\gamma=33.3\times 10^6$ s$^{-1}$, $\Omega=5\gamma$, $\delta=0$, 
$\Delta_x$= 0.12 $\mu$m, $\la x\ra=-1.322\mu$m, and $\la v\ra=3.613$ m/s
($v_R=k_R\hbar/m=16.15$ m/s, and $v_c=k_c\hbar/m=0.28$ m/s). 
} 
\label{pxf24}
\end{figure}

\begin{figure}
{\includegraphics[width=3.3in]{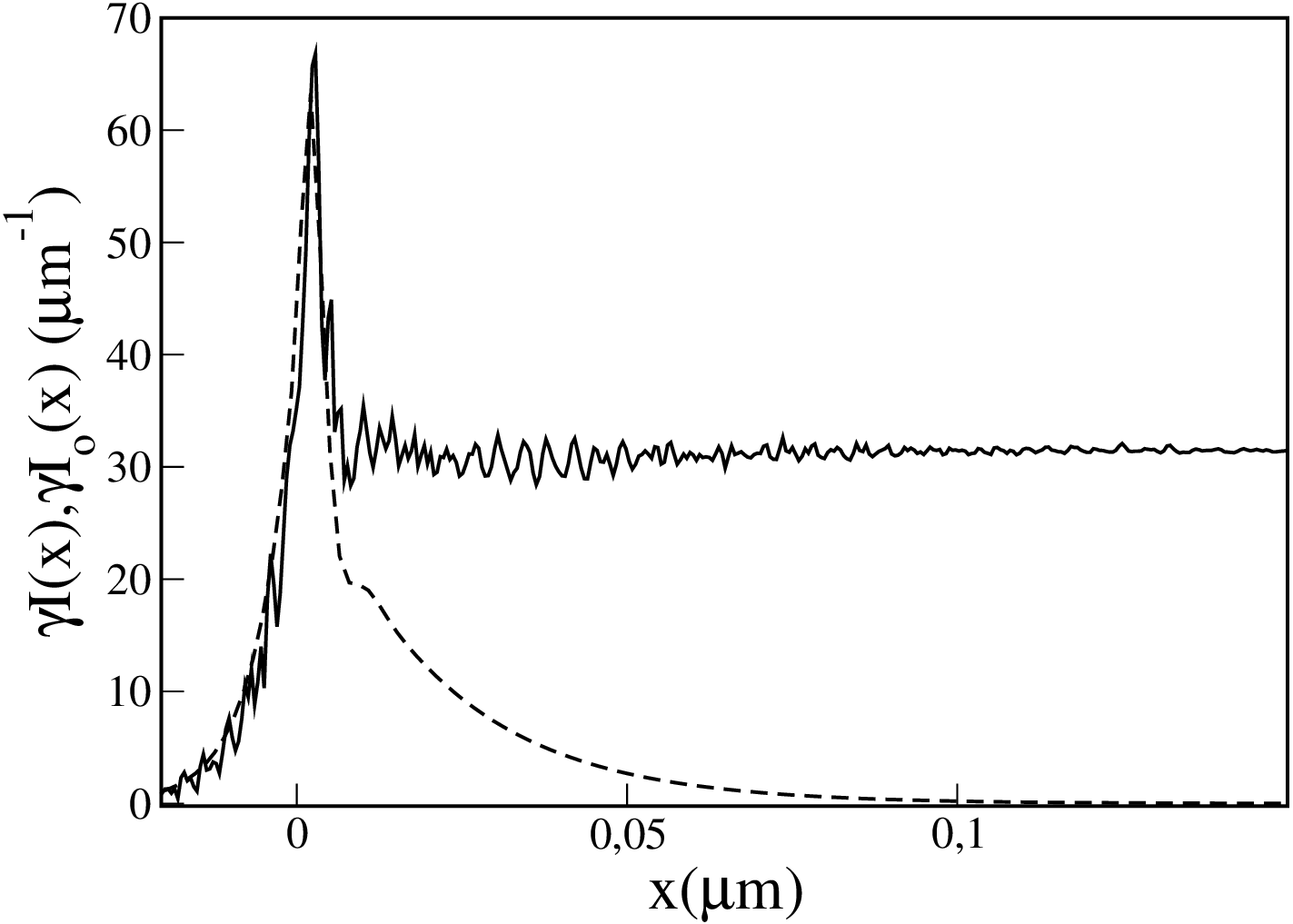}}
\caption{$\gamma I(x)$ (for 40 trajectories, solid line) and
$\gamma I_0(x)$ (dashed line)
versus $x$, same conditions as previous figure except
$\la v\ra=0.18$ m/s. 
}
\label{px23}
\end{figure}

%In terms of these $x$-dependent quantities the Rabi oscillations and
%their suppression  show an interesting variant with respect to the
%temporal analysis that allows to distinguish operationally the
%two  regimes discussed above.   
Above $k_c$, and using the language of the classical 
approximation, the time when the atom arrives at the
laser and starts the Rabi cycle 
is unimportant for an observation of the spatial 
dependence of the photon 
intensity; the only relevant information is the position where 
the emission takes place. In other words, the suppression will not be visible
at the intermediate velocities below $k_R$;
for an example see Figure \ref{pxf24}, where both
$\gamma I_0(x)$ and $\gamma I(x)$ are depicted for the same initial
conditions, with no suppression of the oscillations.   
In the opposite ``quantum'' case, $k<k_c$,  
the suppression effect is visible in $I_0(x)$; for an
example see the dashed line of Figure \ref{pxf24}.
Note the clear distinction between
the small-$x$ region, where there is still a contribution from
$|\lambda_-\ra$, and the larger-$x$ region, which depends only on
$|\lambda_+\ra$.  A similar result holds for $I(x)$ but, because of
the continuation of pumping-emission  cycles, there is no
exponential decay;
see Figure \ref{px23} for an example. We may estimate $I(x)$ after the
transient peak by approximating     
$\la x|\rho_{22}(t)|x\ra$ in Eq. (\ref{pofx2}) by its value at $\gamma=0$,
as in Section \ref{rospd}.
%using Eqs. (\ref{tde}) and (\ref{klkc}).
This 
gives,  with $k\ll k_c$,  $I(x)\approx \la v\ra/v_c^2$. The exponential 
decay of $I_0(x)$ may also be 
approximated in the strong driving limit and $k\ll k_c$ by retaining
the dominant  
exponentially decaying terms in Eq. (\ref{klkc}),  
$I_0(x)\approx \la v\ra e^{-\gamma(m/\Omega\hbar)^{1/2}/2}/v_c^2$.

%Figure 

%Figure 

\section{Discussion\label{disc}}
The possibility to reach ultra cold temperatures for atomic motion 
leads to  a number of new interesting quantum phenomena that 
would remain hidden otherwise. As is well known, the internal level
populations  of fast atoms exhibit  Rabi oscillations when traversing a
sufficiently strong traveling laser field. In this paper, however, we
have shown that for cold atoms these Rabi oscillations may be
suppressed and in fact can be totally absent. In the context of
  moving two-level atoms, two types of Rabi oscillations can occur, temporal
and spatial ones. In the temporal case one observes the population of
the upper level in time and the associated photon emission, regardless
of the atomic position. In the spatial case one observes where the
emissions occur in space, without regard to time.

We have distinguished two regimes for  the suppression of  
temporal Rabi oscillations 
at low and intermediate kinetic energies and have characterized the 
transitions quantitatively. Spatial oscillations are only suppressed at 
low kinetic energies. 

Quite generally the effects of slow atomic 
motion or laser displacement on a two-level qubit 
are of interest in the design
on quantum computers \cite{SH03}. 
Applications of the present effect may be based on the fact that 
at low kinetic energy the  suppression of Rabi oscillations is  
associated with the projection onto specific internal pure states, which 
can be controlled via detuning. Purification procedures 
are essential in quantum data processing \cite{CEM99}.
Our results are also relevant for the 
analysis of time scales defining the particle's traversal through
a region of space \cite{tqm}. They suggest that the excited state population 
cannot in general be taken  as a measure of the time elapsed in the
laser region 
(based solely on a comparison with the Rabi oscillation of the atom at
rest) unless the particle's kinetic energy is sufficiently high. 

Even though we have centered the discussion on an atom-laser system,
the formalism is
equally valid for other two-level systems, and in particular for 
a localized magnetic field acting on a spin-$1/2$ particle,
i. e., to the so called ``Larmor-clock'', which is usually applied with an 
additional potential barrier and in the weak coupling limit \cite{tqm}.    
    
A simplification of our treatment has been the 
semi-infinite laser-illuminated region. This is not a major obstacle
to observe  the effect since, due to the low velocities implied,   
actual finite laser profiles  would be enough for the occurrence of 
many Rabi oscillations of a classically moving atom. 
For example, a Rabi period for our Cs 
transition takes 0.036 $\mu$s when $\Omega=5\gamma$, whereas 
an atom with speed ten times $v_c$ would only move 0.1 $\mu$m in that period.

\begin{acknowledgments}
We are grateful to R. F. Snider for very useful comments.  
This work has been supported
by Ministerio de Ciencia y Tecnolog\'\i a (BFM2000-0816-C03-03), 
UPV-EHU (00039.310-13507/2001),
and the Basque Government (PI-1999-28).
\end{acknowledgments}

\appendix

\section{Limits of the 1D model\label{limits}}
A simple random, classical model will enable us 
to estimate a maximum time of validity 
of the one dimensional approximation used in the text.
At variance with other random walk approaches dealing with 
stimulated emission and absorption processes
in standing waves \cite{ABS81},
the present model 
focuses on the momentum kicks due to spontaneous photon emissions.

Suppose that the atom suffers random kicks at $j=1,2,3,...,n$ 
instants separated by the interval $\Delta t$ and that at  
$t=0$ the atom is at rest, $y=v=0$, where $v$ is the velocity
in the y-direction.  
If $\Delta v_j$ and $\Delta y_j$ are velocity and position increments in
the $y$-direction in step $j$
and $v_y$ and $y_j$ the corresponding velocity and position, 
we have 
\beq
y_n=\sum_j \Delta y_j=\Delta t\sum_j v_j
\eeq
and  
\beqa
\sum_j v_j&=&
(\Delta v_1)+(\Delta v_1+\Delta v_2)+...+(\Delta v_1+...+\Delta v_n)
\nonumber\\
&=&n \Delta v_1+(n-1)\Delta v_2+(n-2)\Delta v_3+....+\Delta v_n,
\nonumber\eeqa
so that 
\beqa
y_n^2&=&(\Delta t)^2[n^2(\Delta v_1)^2+
(n-1)^2(\Delta v_2)^2+...+(\Delta v_n)^2]
\nonumber \\
&+& {\rm crossed\; terms\; involving\;} \Delta v_i\Delta v_l, i\ne l.
\nonumber
\eeqa
When taking the average, the contribution from the crossed terms vanishes, 
\beqa
\la y_n^2\ra&=&(\Delta t)^2[n^2+(n-1)^2+...+1]\la (\Delta v)^2\ra
\nonumber
\\
&=&\frac{n(n+1)(2n+1)}{6}
(\Delta t)^2\la (\Delta v)^2\ra,
\nonumber
\eeqa
where $\la (\Delta v)^2\ra $ is the variance of the
velocity increments in the $y$ 
direction at any of the $j$ instants.
For large $n$,  
\beq
\label{y2}
\la y^2\ra\approx \frac{n^3}{3}(\Delta t)^2\la (\Delta v)^2\ra.  
\eeq
For the dipole radiation distribution of Eq. (\ref{dis}), and
considering that the modulus of each 
velocity increment is due to the recoil velocity
$\Delta v=k_L\hbar/m$, we find 
$\la (\Delta v_y)^2\ra=2\Delta v^2/5$. 
A critical $n$ may be obtained from (\ref{y2}) by imposing, say,
$k_L\la y^2\ra^{1/2}=1/10$.
For the transition of Cs atoms we are considering, 
$\gamma=33.3\times 10^6$ s$^{-1}$,
$k_L^{-1}=852/2\pi$ nm,  and for strong driving
$\Delta t\approx 2/\gamma$;  this gives 
$n\approx 30$, or approximately 2 $\mu s$ of time from 
the first photon emission.

The other phenomenon to take into account is quantum
mechanical dispersion of the 
wave packet. This may be estimated by assuming
$k_L\la y^2\ra^{1/2}=1/10$
in the formula
\beq
(\Delta y)^2=(\Delta y)_0^2
\left[1+\frac{\hbar^2t^2}{4m^2(\Delta y)_0^4}\right].
\eeq
For $k_L(\Delta y)_0=1/20$, this gives 3 $\mu$s.

\section{Theory with decay\label{decay}}

We provide generalizations of Eqs. (\ref{A14}-\ref{d}) for 
$\gamma\ne 0$ and an arbitrary value of $y$ corresponding to the 
Hamiltonian of Eq. (\ref{hamga}): 
\beq
\lambda_\pm=-\frac{1}{2}\Bigg[\delta+i\frac{\gamma}{2}
\pm\Bigg(\delta^2-\frac{\gamma^2}{4}+i\delta\gamma
+\Omega^2\Bigg)^{1/2}\Bigg],
\eeq
\beq
|\lambda_{\pm}\ra=\left({1}\atop{\frac{2\lambda_\pm}{\Omega
 e^{-ik_Ly}}}\right),
\label{eigenv}
\eeq
\beq
k_\pm^2=k^2-\frac{2m\lambda_\pm}{\hbar},\;
q^2=k^2+\frac{2m}{\hbar}(\delta+i\gamma/2),
\eeq
where $k_\pm$ and $q$ have positive imaginary parts. In terms of these 
variables, the expressions for $C_\pm$, $R_1$, and $D$ are the same as in 
Eqs. (\ref{A18}) and (\ref{d}), whereas  
\beq
R_2={k(k_+-k_-)\Omega}e^{ik_Ly}/{D},
\eeq
In this one-dimensional approximation the parametric dependence 
on $y$ does not affect 
the reflection probabilities, $\lambda_\pm$, $q$, $k_\pm$, 
or $C_\pm$, but it gives a phase factor to $R_2$,
and to $\la 2|\lambda_\pm\ra$. A consequence of the later is the 
formation of different pure states in the laser region at low kinetic
energy. In particular, for 
$\gamma=0$,
\beq\label{decayf}
|\lambda_+\ra=\left(
{{1}\atop{-\frac{\delta+\Omega'}{\Omega e^{-ik_Ly}}}}\right).
\eeq

\section{Relation with the Raman-Nath approximation \label{RN}}
It is interesting to compare the approximation of the text 
($x$ as a variable, $y$ as a parameter), with the Raman-Nath 
(RN) approximation where $y$ is the variable, and $x$ a parameter,
expressed in terms of time through an assumed (classical) linear
relation  
\beq\label{time}
t=mx/k\hbar, 
\eeq  
where $k$ is also a parameter. 
Consider the atom evolving with the Hamiltonian
\beq
H_{RN}= \frac{\hbar}{2}
\left({0\atop \Omega e^{ik_Ly}}
{\Omega e^{-ik_Ly}
\atop -2\delta} \right).
\label{hamRN}
\eeq
Note the absence of a kinetic term, a basic feature of the RN
approximation. This Hamiltonian  is easily diagonalized.
The eigenvalues are given by Eq. (\ref{A13})
and the eigenvectors take the form of Eq. (\ref{eigenv}) so that the 
wave vector for 
a state initially in the ground state with amplitude $\psi_0(y)$
is given by 
\beqa
\nonumber
\psi^{(1)}_{RN}(y)&=&{\psi_0}(y)\left[\frac{\Omega'-\delta}{2\Omega'}
e^{-i\lambda_+t}
+\frac{\Omega'+\delta}{2\Omega'}e^{i\lambda_-t}\right],
\\
\nonumber
\psi^{(2)}_{RN}(y)&=&\psi_0(y)\left[\frac{-\Omega e^{ik_Ly}}{2\Omega'}
e^{-i\lambda_+t}+\frac{\Omega e^{ik_L y}}{2\Omega'}
e^{-i\lambda_- t}\right],
\eeqa
or ${\bf \Psi}_{RN}=\psi_0{\bf \Phi}_{RN}$. Making
the substitution of Eq. (\ref{time}) in ${\bf \Phi}_{RN}$ one finds the 
same result as $\bf{\Phi}_k$ of Eq. (\ref{A17}) 
when using the ``high kinetic energy'' approximations of Eqs. (\ref{hek},\ref{heq}),
except
for the plane wave factor $e^{ikx}/(2\pi)^{1/2}$, 
\beq
{\bf \Phi}_k\to \frac{e^{ikx}}{(2\pi)^{1/2}}{\bf \Phi}_{RN}.
\eeq
Physically, the plane wave accounts for the undisturbed
translational motion
along the incident direction  whereas 
the internal one is described by 
the RN wave vector ${\bf \Phi}_{RN}$.

In the RN approximation the probability to find the
atom in the excited state 
is given by $\Omega^2\sin^2(t\Omega'/2)/\Omega'^2$. 
In spite of the neglect of the spatial displacement along $y$, 
there is a possibility of momentum exchange due to the absorption of a
laser photon \cite{TRC84}. 
Taking the Fourier transform we find the following normalized
momentum distributions
for ground and excited state
\beqa
{\cal P}_1(p_y)&=&|\la p_y|\psi_0\ra|^2,
\\
{\cal P}_2(p_y)&=&|\la p_y-k_L\hbar|\psi_0\ra|^2.
\eeqa
Note that the RN approximation cannot describe the Rabi-oscillation-suppression 
effect at
low kinetic energies 
or the transition region around $k_c$ since the longitudinal motion is
treated classically. The suppression effect at intermediate 
energies could be described if an ensemble of ``RN atoms''  
mimicking a wave packet distributed along the longitudinal direction
is considered.


\begin{thebibliography}{10}
\bibitem{Sacchetti01} A. Sacchetti, J. Phys. A {\bf 34}, 10293 (2001). 
\bibitem{EFYS02} M. A. Efremov, M. V. Fedorov,
V. P. Yakovlev, and W. P. Schleich, quant-ph/0209134
\bibitem{Hegerfeldt91}  G.~C. Hegerfeldt and T.~S. Wilser, in:
{\it Classical and Quantum Systems.} 
Proceedings of the Second International Wigner Symposium, July
1991, edited by H.~D. Doebner, W. Scherer, and F. Schroeck, (World
Scientific, Singapore, 1992), p. 104;
G.~C. Hegerfeldt,
\newblock Phys. Rev. A {\bf 47}, 449 (1993); G.~C. Hegerfeldt and
D.G. Sondermann, Quantum 
  Semiclass.~Opt.~{\bf 8}, 121 (1996). For a review cf.  M.~B. Plenio
  and P.~L. Knight, 
Rev. Mod. Phys. {\bf 70}, 101 (1998). The quantum jump approach is
essentially equivalent to the Monte-Carlo wave function approach of 
J. Dalibard, Y. Castin and  K. M{\o}lmer,  
    Phys. Rev. Lett., {\bf68}, 580 (1992), and to the quantum trajectories of 
 H. Carmichael, {\em An Open Systems Approach to Quantum 
Optics}, Lecture Notes in Physics m18, (Springer, Berlin,  1993).
\bibitem{Cohen-Dalibard85} C. Cohen-Tannoudji and J. Dalibard, 
Eurphysics Letters {\bf 1}, 441 (1986). 
\bibitem{foot1}$I(x)$ 
may be also 
interpreted as a dwell time density for the
excited state at $x$ \cite{tqm}.
\bibitem{tqm} J. G. Muga, R. Sala and I. L. Egusquiza (eds.), 
{\it Time in Quantum Mechanics} (Springer, Berlin, 2002).
\bibitem{DEHM02} J. A. Damborenea, I. L. Egusquiza, G. C. Hegerfeldt, and 
J. G. Muga, Phys. Rev. A {\bf 66}, 052104 (2002). 
\bibitem{WM} D. F. Walls and G. J. Milburn, {\it Quantum Optics}, 
(Springer, Berlin, 1994), Ch. 11. 
\bibitem{foot2}A $\gamma$-corrected $k_c$ is defined as  
the value of $k$ that makes zero the real part of $k_-^2$, 
that is, $k_c=[\frac{m}{2}(4\Omega^2-\gamma^2)^{1/2}]^{1/2}$.
A numerical example 
may be seen in 
in Figure \ref{swanalitic1}.
\bibitem{Hegerfeldt93} G. C. Hegerfeldt, Phys. Rev. A {\bf 47}, 449 (1993). 
% How to reset an atom
\bibitem{Hensinger01} W. K. Hensinger at al., Phys. Rev. {\bf 64},
033407 (2001). 
\bibitem{SH03}S. Shresta and B. L. Hu, quant-ph/0301180,
and references therein.
\bibitem{CEM99} J. I. Cirac, A. K. Ekert, and C. Macchiavello,
Phys. Rev. Lett. {\bf 82}, 4344 (1999).
\bibitem{ABS81}E. Arimondo, A. Bambini, and S. Stenholm, 
Phys. Rev. A, {\bf 24}, 898 (1981).  
\bibitem{TRC84} C. Tanguy, S. Raynaud, and C. Cohen-Tannoudji, 
J. Phys. B: At. Mol. Phys. {\bf 17}, 4623 (1984). 
%\bibitem{KK97} A. R. Kolovsky and H. J. Korsch, Phys. Rev. A
%{\bf 55}, 4433 (1997). 
%\bibitem{MF02} C. G. Meister and H. Friedrich,
%Phys. Rev. A {\bf 66}, 042718 (2002). 

%\bibitem{SPBCM92} T. Sleator, T. Pfau, V. Balykin, O. Carnal, and 
%J. Mlynek, Phys. Rev. Lett. {\bf 68}, 1996 (1992). 
\end{thebibliography}
\end{document}